\documentclass[a4paper,11pt,hyper]{JHEP3}
\usepackage{amsfonts,latexsym,graphicx,epsfig,amssymb,mathrsfs,amsmath}

\newcommand{\bm}[1]{\hbox{\boldmath{$#1$}}}

\newcommand{\dd}{{\rm d}}

\newcommand{\phN}{\partial_{\cal N} \phi}

\newcommand{\PP}{{\cal P}}
\newcommand{\BB}{{\cal B}}

\title{$\delta N$ formalism from superpotential and holography}
\author{
Jaume Garriga$^{a,b}$, Yuko Urakawa$^{c,d}$, Filippo Vernizzi$^{e,f}$
\\
a. Departament de F{\'\i}sica Fonamental i Institut de Ci{\`e}ncies del Cosmos, \\
Universitat de Barcelona, Mart{\'\i}\ i Franqu{\`e}s 1, 08028 Barcelona, Spain\\
b.  Institute of Cosmology, Department of Physics and Astronomy,\\
Tufts University, Medford, MA 02155, USA\\
c. Department of Physics and Astrophysics, Nagoya University,\\ Chikusa,
Nagoya 464-8602, Japan \\
d. School of Natural Sciences, Institute for Advanced Study, \\ Olden Lane, 
Princeton, NJ 08540, USA \\
e.  Institut de physique th\' eorique, Universit\'e  Paris Saclay, \\ CEA, CNRS, 91191 Gif-sur-Yvette, France \\
f. Physics Department, Theory Unit, CERN, CH-1211 Gen\`eve 23, Switzerland 
}

\abstract{We consider the superpotential formalism to describe the
evolution of 
$D$
scalar fields during inflation, generalizing it to include
the case with non-canonical kinetic terms. We provide a characterization
of the attractor behaviour of the background evolution in terms of first
and second slow-roll parameters (which need not be small). We find that the superpotential
is useful in justifying the separate universe approximation from the gradient expansion, 
and also in computing the spectra of primordial perturbations around attractor
solutions in the $\delta N$ formalism. As an application, we consider a class of models where the
background trajectories for the inflaton fields are derived from a product separable superpotential.
In the perspective of the holographic inflation scenario,  such models are dual
to a deformed CFT boundary theory,  with $D$ mutually uncorrelated
deformation operators. We compute the bulk power spectra of primordial adiabatic and entropy cosmological perturbations, and show that the results agree with the ones obtained by using conformal perturbation theory in the dual picture. }

\preprint{CERN-PH-TH-2015-231}

\maketitle

\begin{document}

\section{Introduction}

Measurements of the temperature anisotropies and polarization of the cosmic microwave
background place stringent constraints on a wide range of inflationary models.
While current data are  consistent with single-field inflation, multi-field scenarios
arise quite naturally in most attempts to embed  inflation within a broader theory, and it is therefore 
important to address this more generic situation.

The amplitude of primordial perturbations is often described in terms of $\zeta$, 
the curvature perturbation on hypersurfaces of constant energy density. In single-field inflation this quantity is conserved on super-horizon scales,
but in multi-field inflation it  can evolve after Hubble crossing \cite{Wands:2000dp,Lyth:2004gb,Langlois:2005qp}. To take this evolution
into account, it is convenient to use the so-called $\delta
N$ formalism~\cite{Starobinsky:1986fxa, Salopek:1990jq, SS,
Sasaki:1998ug, LMS}, which gives the
correlators of $\zeta$ at the end of inflation in terms of the
correlators of the scalar-fields fluctuations at  Hubble
crossing. The relation between them is established by considering the background evolution of an ensemble of
homogeneous universes with different initial conditions (for a review,
see Refs.~\cite{LR98, Tanaka:2010km}). This approach is not restricted to the correlators of $\zeta$, but can also be applied to
 entropy perturbations (see e.g.~\cite{Sasaki:2006kq,LVW08}).

A useful tool for describing the evolution on super-horizon scales is the Hamilton-Jacobi (H-J) formalism
first introduced by Salopek and Bond~\cite{Salopek:1990jq} (see
also~\cite{Skenderis:2006jq, Skenderis:2006fb}). This was originally developed for
the case of $D$ scalar fields $\phi^I$ 
(here and below capital latin indices $I, J, K, \ldots$ run from $1$ to $D$)
with canonical kinetic terms, but for generality here we shall consider an extension to Lagrangians 
of the form $P(X^{IJ},\phi^K)$, where 
\begin{equation}
X^{IJ}\equiv-(\partial_\mu\phi^I\partial^{\mu}\phi^J)/2 \;. 
\end{equation}
The first step is to encode the background dynamics in a
time independent H-J equation for a ``superpotential'' $W$.
This object is just the Hubble rate expressed as a function 
of the field values $\phi^I$, 
\begin{equation}
W = \frac12 H(\phi^I, c_I)\;,
\end{equation}
where a complete solution of the H-J equation contains an equal number of integration constants 
$c_I$, which account for
the freedom in the initial values of the field momenta.
As we shall see, cosmological evolution can then be seen as a ``gradient flow'' of $W(\phi^I)$ in field space.
The purpose of this paper is to further develop the $\delta N$ formalism by taking advantage of this description.

The superpotential approach is also interesting in connection with the
possibility of a holographic description of inflation
\cite{Larsen:2002et, LM03,vdS,  LM04}. For the case of de Sitter, this
idea was first considered in Refs.~\cite{Strominger,Witten}, by analogy
with the gauge/gravity duality which holds in asymptotically AdS spaces
(see also Refs.~\cite{Bousso:2001mw,Strominger2}). Recently, field
theories which are dual to de Sitter have been identified for the case
of higher spin gravity \cite{Anninos:2011ui}. 
On the other hand, for the case of Einstein gravity, the duality remains
at an exploratory stage. In  the absence of a more concrete realization,
a fruitful strategy has been to focus on small deformations of a generic
boundary conformal field theory (CFT), with Lagrangian of the form
${\cal L} = {\cal L}_{CFT} + \sum_I g^I {\cal O}_I$.  
This setup  is characterized by a few parameters, such as the central
charge of the CFT, and the operator product expansion coefficients 
for the deformation operators ${\cal O}_I$. 
Such parametrization allows for some explicit calculations, which can be done by using conformal perturbation theory ~\cite{BMS,JYcsv,JKY14}. 
With the identification $g^I=\phi^I$ (up to a proportionality constant), 
the renormalisation group (RG) flow of the couplings $g^I$ in the boundary theory corresponds to field evolution in the bulk, while the superpotential $W$ plays the role of a $c$-function for the RG flow.

The case with a single deformation operator ${\cal O}$ corresponds to single field inflation, which has been extensively considered in the 
literature~\cite{Larsen:2002et, LM03,vdS,  LM04, BMS, JYcsv,
Maldacena2002, Seery:2006tq,Maldacena:2011nz, Schalm:2012pi,
Mata:2012bx, JYsingle, Ghosh:2014kba, Larsen:2014wpa, Banks:2013qra,
Banks:2013qpa, Kiritsis:2013gia, Kol:2013msa, Kawai:2014vxa,
Binetruy:2014zya, MS_HC09,MS_HC10, MS_HCob10, MS_NG, MS_NGGW, McFadden:2013ria}.  
In particular, Refs.~\cite{BMS, JYcsv} studied the power spectrum and bispectrum of $\zeta$, showing agreement between the boundary and bulk calculations.\footnote{Furthermore, in Ref.~\cite{JYcsv}, it was shown that the power spectrum of $\zeta$ is conserved at large scales, as expected in
the standard cosmological perturbation theory for the case of single
field models. The conservation of higher order correlation functions,
however, remains an unresolved open issue.}  
The four-point correlation function was also computed in Ref.~\cite{Ghosh:2014kba}, recovering the result from the bulk calculation of Ref.~\cite{Seery:2008ax} in the slow-roll regime.

More recently, Ref.~\cite{JKY14} extended the holographic approach  to multi-field
inflation by considering a CFT with $D$ mutually uncorrelated deformation operators ${\cal O}_I$, and the primordial power spectra for adiabatic and entropy
perturbations were computed in conformal perturbation theory. As an application of the methods presented in this paper, here we will compare the results of Ref.~\cite{JKY14}
with a bulk calculation based on the $\delta N$ technique. 

The paper is organized as follows. In Sec.~\ref{Sec:super} we review the superpotential formalism. We also give a characterization of the attractor behavior of background solutions in terms of the 
first and second slow-roll parameters. Note, however, that such parameters will not be required to be small in our discussion. In Sec.~\ref{Sec:RGcosmology} we review the separate universe approximation, 
on which the $\delta N$ formalism is based. As we shall see, the superpotential will be very useful for computing the primordial spectra, 
particularly in the case when the background is an attractor.  The expressions for the primordial spectra of adiabatic and entropy perturbations in terms
of the superpotential are given in Sec.~\ref{Sec:deltaN}. In Sec.~\ref{Sec:holography}, we  compare these results
with those recently obtained in Ref.~\cite{JKY14} from the holographic point of view. In Sec.~\ref{Sec:Case}, we elaborate on
several explicit models of inflation with a product separable
superpotential, which should be dual to a QFT with uncorrelated
multi-deformation operators.   Finally, we conclude in Sec.~\ref{Sec:conclusion}

\section{Superpotential and background evolution} \label{Sec:super}

Consider a scalar field theory in a $(d+1)$-dimensional spacetime, with an action of the form
\begin{align}
 & S= \int \dd^{d+1} x \sqrt{-g}\, P(X^{IJ}\,, \phi^K)\;. \label{Exp:action}
\end{align}
The corresponding energy momentum tensor is given by
\begin{equation}
T_{\mu\nu} =  P_{IJ}\,\partial_\mu\phi^I\partial_\nu\phi^J+Pg_{\mu\nu}, \label{emt}
\end{equation}
where we use the notation
\begin{equation}
P_{IJ} = {\partial P\over \partial X^{IJ}}\,. \label{pij}
\end{equation}
Since $X^{IJ}$ is symmetric about $I$ and $J$, so is $P_{IJ}$. If the gradients of all fields are aligned in the same time-like direction, the energy momentum tensor (\ref{emt}) has the form of a perfect fluid, with pressure $P$ and energy 
density given by 
\begin{equation}
\rho=2 P_{IJ} X^{IJ}-P.
\end{equation}
The perfect fluid form will be valid for our background solution, where all fields depend only on time.
However, in general the fluid will be imperfect for a perturbed solution.

We assume a flat $(d+1)$-dimensional FRW universe described by the metric $\dd s^2=-\dd t^2+a^2(t) \dd\vec x^2$. The background field equations for $\phi^J(t)$  are given by 
\begin{align}
 & \big( P_{IJ} \dot{\phi}^J \big)\,{\dot{}} + d H P_{IJ} \dot{\phi}^J - (\partial P/ \partial\phi_I ) = 0\,. \label{Eq:KG}
\end{align}
Here 
\begin{equation}
H \equiv {\dot a\over a} 
\end{equation}
is the expansion rate. The Friedmann equation reads
\begin{align}
 & H^2 = \frac{ 2\kappa^2}{d(d-1)} \rho = \frac{ 2\kappa^2}{d(d-1)} \left( 2P_{IJ} X^{IJ} - P \right)\,, \label{Eq:Friedmann}
\end{align}
and
\begin{align}
 & \dot{H} = - \frac{\kappa^2}{d-1} P_{IJ} \dot{\phi}^I \dot{\phi}^J \label{Eq:dH}
\end{align}
where $\kappa^2 \equiv 8 \pi G$ is the gravitational constant.  

As pointed out by Salopek and Bond in Ref.~\cite{Salopek:1990jq} (see also
Refs.~\cite{Skenderis:2006jq, Skenderis:1999mm, DeWolfe:1999cp,
Freedman:2003ax, Kinney:1997ne}), the field equations for canonical
scalar fields can be recast into  first-order form by introducing a superpotential. 
Here, we show that the same treatment applies to non-canonical scalar fields. 
Let us start by defining the momenta
\begin{equation}
\pi_I \equiv {\partial P\over \partial \dot\phi^I}=P_{IJ} \dot{\phi}^J.\label{smallpi}
\end{equation}
In general $P_{IJ}$ can have explicit dependence in field velocities
$\dot\phi^K$, but we assume that our Lagrangian is non-singular, so 
that (\ref{smallpi}) can be solved for the field velocities as a
function of positions and momenta: 
\begin{equation}
\dot\phi^I = F^I(\phi^J,\pi_K). \label{fi1}
\end{equation}
Note that the momenta $\pi_J$ differ from the canonical ones $\Pi_J
=\partial {\cal L}/\partial \dot\phi^J = \sqrt{-g}\ \pi_J$ by a factor
of $a^d$, but with our choice, the relation (\ref{fi1}) does not contain any
explicit time dependence through the scale factor $a$. We may now use
(\ref{fi1}) in order to remove all occurrences of $\dot\phi^I$ in the
energy density. Then, substituting 
\begin{equation}
 H = 2 W(\phi^K)\, , \label{Eq:HW}
\end{equation}
and
\begin{equation}
 \pi_I =P_{IJ} \dot{\phi}^J= - \frac{2(d-1)}{\kappa^2} \frac{\partial
 W}{\partial \phi^I} \,, \label{Eq:dphi}
\end{equation}
which satisfies Eq.~(\ref{Eq:dH}), into the Friedmann equation
(\ref{Eq:Friedmann}), we obtain a partial differential equation for the
superpotential $W(\phi^I)$:
\begin{equation}
W^2 = {\kappa^2\over 2d(d-1)} \rho\left[\phi^K,\partial W/\partial\phi^K\right].\label{HJ}
\end{equation}
Following Ref.~\cite{Salopek:1990jq}, we shall call this the separated, or time independent, 
H-J equation.\footnote{For the case of canonical scalar fields, it is
shown in \cite{Salopek:1990jq} that the separated H-J equation can be
obtained from the Einstein-Hamilton-Jacobi equation, after factoring out
the volume of space in Hamilton's principal function $S \propto a^d W$
.} 
The substitutions (\ref{Eq:HW}) and (\ref{Eq:dphi}) are actually
motivated by the form of the momentum constraint in the long wavelength
limit~\cite{Salopek:1990jq}. We will discuss this constraint in the next
Section, where we consider cosmological perturbations [see
Eq. (\ref{Const:Mgr})]. Nonetheless, for the time being, we may simply
think of  (\ref{Eq:HW}), (\ref{Eq:dphi}) and (\ref{HJ}) as a convenient
reformulation of the background equations of motion. Indeed, in Appendix
\ref{secondorder}, we show that for any solution $W(\phi^K)$ of
(\ref{HJ}), Eqs.~(\ref{Eq:HW}) and (\ref{Eq:dphi}) generate a solution
of the field equations (\ref{Eq:KG}) and (\ref{Eq:Friedmann}).

A complete solution of the H-J equation $W(\phi^K,c_K)$
depending on $D$ independent integration constants $c_K$, can be used
for generating any background solution from the first order equations
(\ref{Eq:dphi}) and (\ref{Eq:HW}). The existence of such a solution can be understood as follows. Given
some initial data $(\phi^K_0,\pi^0_K)$ at $t=t_0$, the background
equations of motion can be solved in order to find
$H(t;\phi^K_0,\pi^0_K)$. If one of the fields has a monotonic
evolution, we can use it as the time variable, and express the initial
positions of the rest in terms of their positions at the time $t$. With
this, we have $H=H(\phi^K;\pi^0_K)$. Finally, we may choose
$c_K=\pi^0_K$ (or any invertible relation between $c_K$ and
$\pi^0_K$), leading to $W=H(\phi^K,c_K)/2$.

\subsection{Cosmological evolution as a gradient flow}

A cosmological solution can be thought of as a trajectory in field space, parametrized by the $e$-folding number $N$. 
For each field $\phi^I$, we may define the corresponding beta function $\beta^I$ as the dimensionless component of the tangent vector $\dd/\dd N $ in 
the corresponding direction, 
\begin{equation}
\frac{\dd}{\dd N}  = \frac{1}{\kappa} \beta^I \frac{\partial}{\partial
 \phi^I} \;, \qquad \beta^I \equiv \kappa \frac{\dd \phi^I}{\dd N} \;. \label{betaup}
\end{equation}
The standard ``first'' slow-roll parameter can then be written in terms of the beta functions as
\begin{equation}
\varepsilon_1 \equiv - \frac{\dot{H}}{H^2} = \frac{\kappa^2 (\rho+P)}{(d-1)H^2}
 = \frac{P_{IJ} \beta^I\beta^J}{(d-1)} \geq 0\;, \label{Def:ep} 
\end{equation}
where the last inequality holds provided that the null energy condition is satisfied. In what follows, we shall assume that $P_{IJ}$ is non-degenerate. 

In that case, $P_{IJ}$ can be thought of as a metric, which we may use in order to raise and lower indices. 
Introducing
\begin{equation}
f\equiv-\ln W \label{cfunction},
\end{equation}
Eq.~\eqref{Eq:dphi} can be written as
\begin{equation}
 \beta_I \equiv  P_{IJ}  \beta^J =  \frac{d-1}{\kappa} \frac{\partial f}{\partial \phi^I}\;, \label{beta_def}
\end{equation}
which expresses the beta functions as a gradient.
From  Eq. (\ref{mono})
\begin{equation}
\frac{\dd f}{\dd N}  = \frac{ \beta^2 }{d-1}  = \varepsilon_1> 0 \;, \label{mono}
\end{equation}
and  therefore $f$ is monotonically increasing along any cosmological
trajectory. Here, we have used the notation $\beta^2\equiv \beta_I\beta^I$. 
Also, since we assume that $P_{IJ}$ is positive definite, the last
inequality is strict provided that at least one of  the fields is
moving. 

If we interpret the $\beta^I$ as beta functions for couplings $g^I=\kappa\phi^I$ in the boundary theory, then $f$ 
plays the role of a $c$-function which decreases monotonically along the RG flow (note that the 
RG flow proceeds from the UV to the IR, which is opposite to the
direction of cosmological evolution as a function of $N$). 
In this interpretation, $P_{IJ}$ corresponds to the Zamolodchicov metric
in the space of couplings \cite{zamo}. Such metric is needed in order to
relate the components of the RG flow tangent vector $\beta^I$, which
have the upper index, to the gradient of a $c$-function, whose
components have the lower index. We shall further elaborate on the dual
picture in Sec.~\ref{Sec:holography}.

\subsection{Attractor behaviour}     \label{attract}

As mentioned above, the integration constants $c_K$ in $W(\phi^K,c_K)$
correspond to the freedom of choosing the initial value of the momenta
$\pi^0_I$ for given initial values of the fields $\phi^K_0$.  
In the present context, the initial conditions for the long wavelength
bulk evolution are implemented around the time of horizon crossing. 
Under a small variation of the integration constants, the gradient flow
will change. It is then important to characterize whether such
dependence remains at late times, or whether it decays.  

First of all, it should be noted that along the field trajectory
\begin{equation}
{\partial W \over \partial c_K} = A^K a^{-d}, \label{decay}
\end{equation}
where $A^K$ are constants and we remind that $d$ is the number of spatial dimensions.
For  the case of scalar fields with canonical kinetic terms, this result was first derived in 
Ref.~\cite{Salopek:1990jq}. We show in Appendix \ref{dilution} that the same result generalizes to Lagrangians with non-canonical kinetic terms.
Eq. (\ref{decay}) tells us that the effect of the integration constants on the Hubble rate ($H=2 W$) decays quite rapidly with the scale factor. Of course, this does not
immediately imply that the gradient flow will always have an attractor behaviour. It is not enough that the Hubble rate converges to the unperturbed value, but 
also the perturbed trajectories should converge to the unperturbed
ones. It seems difficult to formulate a sufficient condition for such
convergence in the general case. Here, we shall consider a necessary
condition, which is also a sufficient condition in the one-field case,
or when there is effectively just an adiabatic perturbation by the end
of inflation. 

The idea is to consider the change in the field momentum, projected on the unperturbed trajectory. The fractional change in $\pi_I\pi^I$ under a small variation $\Delta c_K$ is given by 
\begin{equation}
\Delta \ln \left(\pi_I\pi^I\right) \equiv \delta_1+\delta_2 = {\pi^I\Delta\pi_I + \pi_I\Delta\pi^I \over \pi_J\pi^J}.
\label{deltaf}
\end{equation}
Here, we use the notation $\pi^I= P^{IJ}\pi_J=\dot\phi^I$, and
$\delta_1$ and $\delta_2$ correspond, respectively, to the first and second terms in the right hand side of the last equality.
For the background to be an attractor, we require that these go to zero as the universe expands,
\begin{equation}
\lim_{a\to \infty} \delta_{1,2}= 0. \label{attractor}
\end{equation}
Since the metric $P_{IJ}$ depends on momenta, it will have dependence on
$c_K$, and so the two terms $\delta_1$ and $\delta_2$ can have a
somewhat different behaviour. We require that the condition
(\ref{attractor}) should hold separately for the two terms, so that the
projection of both $\Delta \pi^I$ and $\Delta\pi_I$ on the unperturbed
trajectory can be considered to be small. 

Using Eq. (\ref{Eq:dphi}) it is straightforward to show that 
\begin{equation}
\pi^I \Delta \pi_I = -{4(d-1)\over \kappa^2} W {\dd \over \dd N} \left({\partial W\over \partial c_K}\right) \Delta c_K.\label{delta1}
\end{equation}
 Using (\ref{decay}) in (\ref{delta1}), and (\ref{Eq:Friedmann}), we have
\begin{equation}
\pi^I\Delta \pi_I = \left({\partial_{c_K}}\rho\right)\ \Delta c_K \equiv \Delta\rho. \label{pdp}
\end{equation}
For later convenience, here we have used the Friedmann equation to write
$W^2$ in terms of the energy density $\rho$ (which depends on the $c_K$
only through the kinetic variables $X^{IJ}$). Using (\ref{decay}) again  
in order to evaluate the derivative with respect to $c_K$, we have
\begin{equation}
\delta_1 = {2d \over H\varepsilon_1 a^d}\Delta W_0,\label{deltaf0}
\end{equation}
where $\Delta W_0= A^J \Delta c_J$ is the change in $W$ at the initial
time due to the variation $\Delta c_J$.

From (\ref{deltaf0}) we find that the first condition (\ref{attractor}) for the attractor behavior, concerning 
the behavior of $\delta_1$, will be satisfied provided that
\begin{equation}
 d-\varepsilon_1+\varepsilon_2 > p\geq 0 \label{condition} \;.
 \end{equation}
 Here we have introduced the second slow-roll parameter
\begin{equation}
\varepsilon_2 \equiv \frac{\dd \ln \varepsilon_1}{\dd N} =
 \frac{\dot\varepsilon_1}{\varepsilon_1 H}\,,  \label{Def:etI}
\end{equation}
and a positive constant $p$. Eq.~(\ref{condition}) guarantees that
$\delta_1$ decays faster than $a^{-p}$. If $p$ is small, then the
approach to the attractor can be slow, requiring a time-scale of the
order $\Delta N \sim p^{-1}$ $e$-foldings. An efficient approach to the
attractor on the Hubble time-scale requires 
$p\gtrsim 1$.

In addition, we need to consider the behaviour of $\delta_2$, which may place additional restrictions. Since $\pi_I\pi^I = (\rho+P)$, from (\ref{deltaf}) and (\ref{pdp}) we immediately obtain
\begin{equation}
\pi_I\Delta \pi^I = \left({\partial_{c_K}}P\right) \Delta c_K \equiv\Delta P. \label{pdp2}
\end{equation}
Note that the pressure $P$ only depends on the $c_K$ through the kinetic variables $X^{IJ}$, and therefore 
\begin{equation}
\partial_{c_K}P = P_{IJ}\, \partial_{c_K} X^{IJ}.
\end{equation}
The same is true of the energy density,
\begin{equation}
\partial_{c_K}\rho= \rho_{IJ}\,  \partial_{c_K} X^{IJ},
\end{equation}
where $\rho_{IJ} = P_{IJ} +2 X^{KL} P_{KL,IJ}$ is the symmetrized partial derivative of $\rho$ with respect to $X^{IJ}$. Since in general $\rho_{IJ}\neq P_{IJ}$, $\delta_2$ can behave differently than $\delta_1$.

For instance if the background solution satisfies the relation
\begin{equation}
\Delta P = c_s^2\ \Delta\rho, \label{sof}
\end{equation}
for some function $c_s^2$, then we have 
\begin{equation}
\delta_2 = {\pi_I\Delta \pi^I \over \pi_J \pi^J} = c_s^2\delta_1={ 2c_s^2 d\over H\varepsilon_1 a^d}\Delta W_0.\label{deltaf2}
\end{equation}
The second condition in (\ref{attractor}), concerning the behavior of $\delta_2$ will then be satisfied provided that
\begin{equation}
 d-\varepsilon_1+\varepsilon_2 -s> p\geq 0 \label{condition2} \;.
 \end{equation}
 This differs from (\ref{condition}) by the last term in the left hand side of the inequality, which is defined as
 \begin{equation}
 s\equiv {\dd \ln c_s^2 \over \dd N}. \label{sdef}
 \end{equation}
 Although (\ref{sof}) is not completely general, it does cover some important cases. 
 For instance, if the Lagrangian is linear in $X^{IJ}$, then the metric $P_{IJ}=G_{IJ}(\phi^I)$
 is just  a function of the fields, and (\ref{sof}) is satisfied with the speed of sound $c_s^2=1$. In this case $\delta_1=\delta_2$ and the conditions (\ref{condition}) and (\ref{condition2}) coincide.
 Another interesting example where (\ref{sof}) is satisfied with a
 non-trivial speed of sound $c_s\neq 1$ is the case of a multi-field DBI
 action, which is discussed in \cite{TDSD}~\footnote{More generally, 
when 
$P_{KL, IJ}$ is given by a linear combination of $P_{IJ}$ as 
$$
 P_{KL, IJ} = f_1 P_{IJ} P_{KL} + f_2 (P_{IK} P_{JL} + P_{IL} P_{JK})\,,
$$ 
where $f_1$ and $f_2$ are functionals of $\phi^I$ and $X^{IJ}$,
$\Delta P$ and $\Delta \rho$ are linearly related as in
 Eq.~(\ref{sof}). In such a case, the second attractor condition
can be formulated as \eqref{condition2}. }. Finally, Eq. (\ref{sof})
 also applies to generic one field models. 
 It is easy to check that in
 such case, the attractor condition (\ref{condition2}) is related to the
 absence of a growing mode for the curvature perturbation on uniform
 field hypersurfaces \cite{GM}. 

 For a generic multi-field model the condition that $\delta_2$ vanishes
 in the asymptotic future will be more elaborate than
 (\ref{condition2}), and should be worked out on a case by case basis. 
 We leave this as a subject for further research.

\section{Separate universe approximation and $\delta N$ formalism
}  \label{Sec:RGcosmology}
The approximate homogeneity and isotropy of cosmological evolution entails a useful relation
between the curvature perturbation and the differential $e$-folding
number, which is used in the so-called $\delta N$ formalism. 
Let us start by reviewing the conditions for the realisation of a universe with such approximate symmetries.
The superpotential formalism will be useful in clarifying the conditions for the validity of this approach,
particularly in the implementation of the momentum constraint.  

\subsection{Separate universe approximation}

The $\delta N$ method is based on the ``separate universe
assumption''. This is the statement that when a characteristic (physical)
scale $L$  of fluctuations is much bigger than the Hubble length $H^{-1}$, i.e.~$L \gg H^{-1}$,  each region of the universe of Hubble size  evolves as a locally homogeneous and isotropic FRW universe. This is justified if each Hubble
patch is approximately homogeneous and isotropic up to corrections of
order $\epsilon \ll 1$, with
$$
\epsilon \equiv \frac{1}{LH}\,. 
$$
Then, since different Hubble patches are causally
disconnected for local theories, they should evolve independently of one another.  

Consider the ADM line element, 
\begin{align}
 & \dd s^2 = - \alpha^2 \dd t^2 + \gamma_{ij} (\dd x^i + \beta^i \dd t)
 (\dd x^j + \beta^j \dd t)\,,
\end{align}
where $\alpha$ and $\beta^i$ are the lapse function and the shift
vector, and $\gamma_{ij}$ is the spatial metric. Here, the shift vector
$\beta^i$ with the index $i$ should be distinguished from the beta
function $\beta^I$ with the index $I$. We  parametrize  $\gamma_{ij}$ as
\begin{align}
 & \gamma_{ij} =  a^2 e^{2 {\cal R}(x)} [e^{h (x)} ]_{ij} \;, \qquad {\rm tr} \, h=0 \,.
\end{align}
The time-like congruence orthogonal to $t=const.$ slices has a unit
tangent vector given by $n^\mu = \alpha^{-1} (1,\, - \beta^i)$, and its
expansion $K$ is given by 
\begin{align}
 & K \equiv \nabla_\mu n^\mu = \frac{1}{\sqrt{-g}} \partial_\mu
 (\sqrt{-g} n^\mu)  = \alpha^{-1} \left[ d(H + \dot{{\cal R}}) - D_i \beta^i \right] \,, \label{Def:K}
\end{align}
where $D_i$ is the covariant derivative defined with respect to the
spatial metric $\gamma_{ij}$. The $e$-folding number may be defined as the
integral of the expansion along the normal congruence: 
\footnote{Alternatively, we may define
the co-moving $e$-folding number $ {\cal N}_c \equiv \int (K_c/d)  \dd \tau$.
Here $\dd \tau = \sqrt{\alpha^2- \beta_i \beta^i} \dd t$ is the element of proper time along a co-moving worldline, 
and $K_c$ is the expansion of the co-moving congruence,  whose unit tangent vector is given by $n^\mu_c =(\alpha^2-\beta^i\beta_i)^{-1/2} (1,\vec 0)$. 
It is straightforward to check that for $\partial_i \beta^i={\cal O}(\epsilon^2)$, ${\cal N}_c \approx {\cal N}$, up to terms quadratic in the spatial gradients.}
\begin{align}
 & {\cal N} \equiv \frac{1}{d} \int K\, \alpha\, \dd t  \label{Def:calN}\,.
\end{align}

In what follows, the metric is assumed to take the FRW form in the long wavelength limit $\epsilon\to 0$, and we shall use the gauge where the traceless matrix $h$ is also transverse,
\begin{equation}
\partial^i h_{ij}=0.
\end{equation} 
For scalar fields, the anisotropic stress $T_{ij} -(1/d) \gamma^{kl}T_{kl} \gamma_{ij}$ is of order ${\cal O}(\epsilon^2)$.
As shown in Ref.~\cite{SKF}, using Einstein's equations, after neglecting terms of lower order in $\epsilon$, which decay like $a^{-d}$, we have
\begin{equation}
\partial_j\beta^i = {\cal O}(\epsilon^2), \label{negl} 
\end{equation}
and
\begin{equation}
\dot h_{ij} = {\cal O}(\epsilon^2). \label{doth}
\end{equation} 
Using these conditions and introducing the derivative with respect to the number of $e$-foldings ${\cal N}$,
\begin{align}
 & \partial_{\cal N} \equiv \frac{d}{K \alpha} \partial_t \,, \label{dNtodt}
\end{align}
the Einstein equations and the field
equations of the scalar fields read~\cite{NTS}
\begin{align}
 &  K^2 = 
 {2 \kappa^2\over d (d-1)} 
 \left( P_{IJ} K^2 \phN^I
 \phN^J - d^2 P  \right) + {\cal O}(\epsilon^2) \,, \label{Const:Hgr}\\
 & \partial_{\cal N} K = - 
 \frac{\kappa^2}{d-1} 
 K P_{IJ} \phN^I
 \phN^J + {\cal O}(\epsilon^2)\,, \label{Eeqgr} \\
 & K \partial_{\cal N} \left( P_{IJ} K \phN^J \right) + d K^2 P_{IJ}
 \phN^J - d^2  (\partial P/\partial \phi^I) = {\cal O}(\epsilon^2)\,. \label{feqgr}
\end{align}
At the leading order in the gradient expansion, these equations coincide
with the background field equations, where $K$ should be understood as
$dH$ for this comparison. For definiteness, in what follows we shall refer 
to Eqs.~(\ref{Const:Hgr})-(\ref{feqgr}), together with Eqs.~(\ref{negl}) and (\ref{doth}),
as the separate universe approximation.

\subsection{Momentum constraint}

The momentum constraint has no counterpart in the background field equations, and so it might enforce
additional requirements for the validity of the separate universe approximation. The momentum constraint is given by
\cite{NTS}
\begin{align}
 &  \partial_i K = - 
 \frac{\kappa^2}{d-1} 
 K P_{IJ} \phN^I \partial_i \phi^J + {\cal
 O}(a\epsilon^3) \,. \label{Const:Mgr}
\end{align}
A factor of $a$ is inserted in ${\cal O}(a\epsilon^3)$, since here we are considering the
comoving derivative of the extrinsic curvature, while 
in our conventions a factor of $\epsilon$ corresponds to a physical gradient.
On the other hand, taking the spatial derivative of the Hamiltonian
constraint (\ref{Const:Hgr}) and using Eqs.~(\ref{Eeqgr}) and
(\ref{feqgr}), we obtain
\begin{align}
 &\partial_i K = - 
 \frac{\kappa^2}{d-1} 
K P_{IJ} \phN^I \partial_i
 \phi^J +  B_i + {\cal O}(a \epsilon^3), \label{dhc}
\end{align}
where $B_i$ is defined as
\begin{align}
  B_i & \equiv 
  \frac{\kappa^2 K }{ (d-1) \partial_{\cal N}  \ln (e^{d {\cal N}} K)} 
 \left[
 \phN^I \partial_i \left(P_{IJ} \phN^J
 \right) - \partial_{\cal N} \left(P_{IJ} \phN^J \right) \partial_i \phi^I \right]\,. \label{Bi}
\end{align}
Here we used Eq.~(\ref{Eeqgr}) to rewrite the denominator on the right hand side.
Comparing (\ref{Const:Mgr}) and (\ref{dhc}) we see that the consistency
of the Hamiltonian and momentum constraint requires that 
\begin{equation}
a^{-1}B_i = O(\epsilon^3). \label{consis}
\end{equation}
Sugiyama, Komatsu and Futamase~\cite{SKF} pointed out that the condition
(\ref{consis}) is automatically satisfied under the slow-roll
approximation. Here, we argue that this conclusion is not restricted to
slow-roll, but follows more generally from the attractor behaviour 
discussed in Subsection \ref{attract}. Indeed, repeating the same argument as in the background, we can express the
field equations at the leading order of the gradient expansion with the
use of the superpotential as
\begin{align}
 & P_{IJ} \phN^J = - \frac{d-1}{\kappa^2} \frac{\partial \ln W}{\partial
 \phi^I}  + {\cal O}(\epsilon^2)\,,  \label{soup}
\end{align}
where the superpotential is now related to the extrinsic curvature by $K=2d W + {\cal O}(\epsilon^2)$. 
In the attractor regime, the dependence of the superpotential $W(\phi^K, c_K)$ on the integration constants $c_K$ can be 
neglected. Then, substituting Eq.~(\ref{soup}) in Eq.~(\ref{Bi}), the leading terms in the two expressions within round brackets in the left hand side of
Eq.~(\ref{Bi}) cancel each other, and we are left with  $a^{-1}B_i = {\cal O}(\epsilon^3)$.

Beyond the attractor regime, we need to consider the dependence of $W$ in the constants $c_K$. In that case, substituting Eq.~(\ref{soup}) in Eq.~(\ref{Bi})
we have
\begin{equation}
a^{-1} B_i = -
{\frac{2 d \, W }{ \partial_{\cal N}  \ln (e^{d {\cal N}} W)}}
{\dd\over \dd{\cal N}}\left( {\partial \ln W\over \partial c_J}\right)( a^{-1} \partial_i c_J) + O(\epsilon^3),
\end{equation}
where the total derivative with respect to ${\cal N}$ is taken along the dynamical trajectories at fixed $c_K$. From Eq.~(\ref{decay}), 
\begin{equation}
{\partial W\over \partial c_J} = A^J e^{-d\cal N} + O(\epsilon^2),
\end{equation}
and after some simple algebra, we have
\begin{equation}
a^{-1}{B_i} = 2 d
\, a^{-d} A^J \, \frac{\partial_i c_J}{a} + O(\epsilon^3). \label{offence}
\end{equation}
The first term in the right hand side contains only one spatial derivative, and so it is naively of order $\epsilon$ in the gradient expansion,  rather than $\epsilon^3$. 
Thus, unless the $c_K$ are constant in space, it might seem that the momentum constraint (\ref{Const:Mgr}) is inconsistent with the spatial derivative 
of the Hamiltonian constraint given in Eq.~(\ref{dhc})\footnote{For instance, it was stated in \cite{Salopek:1990jq} that in single-field models the 
integration constant in the solution of the H-J equation should be constant in space for consistency between the Hamiltonian and momentum constraint. 
But as we argue here, this is not necessarily the case.}.
Note, however, that the term of order $\epsilon$ is accompanied by a decaying function which scales as\footnote{For the case of canonical fields, such scale dependence of the leading
term in the gradient expansion of $B_i$ was first derived in
Ref.~\cite{SKF}.} $a^{-d}$, while the terms of order $\epsilon^3$
include gradients of non-decaying contributions. In particular, for spatial dimension $d>2$, the first term in (\ref{offence})  falls off with physical wavelength faster than $\epsilon^3$, and hence it can be safely ignored for present purposes. For any given co-moving scale, the initial conditions for the long wavelength evolution are generated at horizon crossing, and the first term in (\ref{offence}) 
will be negligible soon after that.

As noted in Ref.~\cite{SKF}, the momentum constraint can always be satisfied in the gradient expansion by modifying (\ref{negl}), so that $\beta^i$ includes terms of lower order 
in $\epsilon$, starting at $a \beta^i = O(\epsilon^{-1})$. As mentioned before Eq.~(\ref{negl}), such lower order terms can be shown to decay as $a^{-d}$ for matter whose anisotropic stress 
is of order $\epsilon^2$. Using the Hamiltonian and momentum constraints at the linearized order in the flat slicing we can obtain the relation~\cite{SKF}~\footnote{When we include the second term in Eq.~(\ref{relNt}), our
equation (\ref{Exp:beta}) differs from the equation (49) in Ref.~\cite{SKF}.}
\begin{align}
 &\partial_i \partial_j \beta^j + B_i =0 \,, \label{Exp:beta}
\end{align}
where we used that 
\begin{align}
 & \partial_t = \left( H - \frac{1}{d} \partial_i \beta^i \right)
 \partial_{\cal N} \,, \label{relNt}
\end{align}
which follows from Eqs.~\eqref{Def:K} and \eqref{dNtodt}, and assuming a flat gauge where ${\cal R} =0$.
Using Eq.~(\ref{Exp:beta}), we can relate 
the term of order $\epsilon^{-1}$ in the expansion of the shift vector $\beta^i$ to the
first term on the right hand side of (\ref{offence}): 
\begin{equation}
\partial_i\beta^i = -{2d\over a^d} \Delta W_0 +O(\epsilon^2),\label{pibi}
\end{equation}
where $\Delta W_0 \equiv A^J\Delta c_J$ is a slowly varying function of position.
After this mode decays,  $a^{-1}B_i$ is formally of order $\epsilon^3$ and
$\partial_i \beta^j = O(\epsilon^2)$, consistent with the separate universe approximation (\ref{negl}).

\subsection{$\delta N$ formalism}

In the separate universe approximation, the last term in square brackets of
Eq.~(\ref{Def:K}) vanishes 
as 
\begin{equation}
D_i\beta^i = {\cal O}\left(\epsilon^2\right)
\end{equation}
in the long wavelength limit.
Inserting Eq.~(\ref{Def:K}) into Eq.~(\ref{Def:calN}) and neglecting
${\cal O}(\epsilon^2)$ 
terms
we obtain  
\begin{align}
 & \delta N(t_2,\, t_1; \bm{x}) \equiv {\cal N} (t_2,\, t_1; \bm{x}) -
 N(t_2,\, t_1) \simeq {\cal R}(t_2, \bm{x}) - {\cal R}(t_1,\,
 \bm{x})\,, \label{long}
\end{align}
where $N$ is the $e$-folding number for the unperturbed background evolution. 

With the help of Eq.~(\ref{long}), one can map the spatial distribution of the scalar fields 
near the time when the relevant scale crosses the horizon, 
$\delta\phi^I_*(\bm{x})$, to the curvature perturbation
at some final space-like hypersurface $\Sigma_e$ near the end of inflation,
${\cal R}(t_e,\bm{x})$. The spatial distribution of the scalar fields should be specified on the
``initial'' space-like hypersurface $\Sigma_*$ where the spatial
curvature vanishes,  
\begin{equation}
{\cal R}(t_*,\bm x)=0.\label{flatslice}
\end{equation}
For the final hypersurface $\Sigma_e$ we have some freedom. For
instance, we may take it to be a hypersurface of constant energy
density,  
\begin{equation}
\rho(t_{\rho_e},{\bm x})=\rho_e=const. \label{constdslice}  
\end{equation}
Note that Eqs.~(\ref{flatslice}) and (\ref{constdslice}) define the
times $t_*({\bm x})$ and $t_{\rho_e}({\bm x})$ as implicit functions of
position. Alternatively, by a suitable choice of coordinates the
hypersurfaces $\Sigma_*$ and $\Sigma_{e}$ can be made to coincide with  
initial and final $t=const.$ slices.
The adiabatic curvature perturbation is defined as:
\begin{equation}
\zeta(t_{\rho_e},\bm{x}) \equiv {\cal R}(t_{\rho_e},\bm{x}). \label{zadi}
\end{equation}
Using $t_1=t_*$ and $t_2=t_{\rho_e}$ in Eq.~(\ref{long}), from
(\ref{flatslice}) and (\ref{zadi}) we obtain the familiar relation
between $\zeta$ and the differential $e$-folding number 
\begin{equation}
\zeta(t_{\rho_e},\bm{x}) = \delta N(t_{\rho_e},t_*;\bm{x}) \equiv \delta
 N(t_{\rho_e}; \delta \phi^J_*(\bm{x})). \label{zetadn} 
\end{equation}
For the last equality we used the separate universe approximation and assumed that the
inhomogeneity in the $e$-folding number can be determined from the initial distribution of the fields
$\delta\phi^J_*(\bm{x})=\delta\phi^J(t_*,\, \bm{x})$. More precisely, the $e$-folding number from the initial to the final hypersurfaces should be computed by solving the 
Friedmann equation in each Hubble patch. In principle, this requires
$(\phi^J_*, \pi_{J*})$ as initial conditions, so the initial time
derivative of the field distribution is also needed. However, as discussed in the previous section, when the
trajectory is an attractor, the dependence on the initial time
derivative dies off rapidly and then the initial distribution can
be expressed only in terms of $\phi^I_*$.  For the one field case, this
is discussed in detail in the following subsection. 

If the inflationary trajectories in field space converge to a unique one \cite{SS, Sasaki:1998ug}, then the adiabatic curvature perturbation $\zeta$ becomes 
subsequently constant in time, {\it i.e.} independent of the value of
the density $\rho_e$ which defines the final hypersurface. However, 
in general, we also need to deal with entropy modes at the final hypersurface. These can be  calculated along similar lines as $\zeta$.

In particular, choosing the uniform field slicing with
$\phi^I=\phi^I_e=const.$ as the final hypersurface, which we may denote
by $\Sigma_{\phi^I_e} (\phi_e^K)$, one obtains a set of $D$ independent
gauge invariant variables $\zeta^{(I)}$ which represent the curvature
perturbation in the different $\delta\phi^I=0$ slicings (see for
instance \cite{Sasaki:2006kq,LVW08}): 
\begin{align}
 & \zeta^{(I)}(\phi^K_e,\, \bm{x})  \equiv {\cal R}(\Sigma_{\phi^I_e},\bm{x}) = \delta N(\Sigma_{\phi^I_e},\, \Sigma_*; \bm{x}) \equiv \delta N^{(I)}(\phi^K_e; \delta \phi^J_*(\bm{x})). 
 \label{zetaI}
\end{align}
Expanding $\zeta^{(I)}(t_e,\, \bm{x})$ in terms of the scalar field
fluctuations  $\delta\phi^J_*(\bm{x})$, we obtain
\begin{align}
 & \zeta^{(I)}(t_e,\, \bm{x}) = \sum_{n=1}^\infty {1\over n!} N_{J_1...J_n}^{(I)} \delta \phi^{J_1}_*(\bm{x})\cdot\cdot\cdot \delta \phi^{J_n}_*(\bm{x}),
 \label{Eq:deltaN}
\end{align}
with
\begin{equation}
N^{(I)}_{J_1...J_n} \equiv \frac{\partial^n N^{(I)}}{\partial \phi^{J_1}_*...
 \partial \phi^{J_n}_*} .
\end{equation}
Such partial derivatives of $N^{(I)}$ can be determined from the solution of the 
equation of motion for the local FRW universe as a function of initial field values 
$\phi^J_*$ in the vicinity of the background solution. 

The relative entropy perturbation $S^{IJ}$ is defined as the difference between two $\zeta^{(I)}$s (see e.g.~\cite{LVW08,Langlois:2011zz}), 
\begin{align}
 & S^{IJ} (t,\, \bm{x}) \equiv d \left[  \zeta^{(I)}  (t,\, \bm{x}) - \zeta^{(J)}(t,\,
 \bm{x}) \right]\,,  \label{Def:SIJ}
\end{align}
which can of course be expanded in powers of $\delta \phi_*^I$ by
substituting Eq.(\ref{Eq:deltaN}).

Finally, we note that the uniform Hubble (or energy density) slicing is given by
\begin{align}
 & 0= \frac{\delta H}{H} = 
 \sum_{I=1}^D {\partial \ln W\over \partial\phi^I}  \delta \phi^I =- \frac{\kappa}{d-1}
 \sum_{I=1}^D \beta_I \delta \phi^I\,.
\end{align}
On the other hand,
at linear order,  the curvature perturbation ${\cal R}$ changes under the time shift 
$t \to t + \delta t$ as ${\cal R} \to {\cal R} - H \delta t$, while the field fluctuations $\delta \phi^I$ change as $\delta \phi^I \to \delta \phi^I - (\beta^I/\kappa) H \delta t$. 
Using these transformations and Eqs.~\eqref{zadi} and \eqref{zetaI}, we have
\begin{align}
\zeta  = {\cal R} - \kappa \sum_{I=1}^D \frac{\beta_I}{\beta^2} \delta \phi^I  = \sum_{I=1}^D { \beta_I \beta^I \over \beta^2} 
 \zeta^{(I)}  \;, 
 \qquad \zeta^{(I)}  = {\cal R} - \kappa \frac{\delta \phi^I}{\beta^{I}}  \;. \label{Rel:zetazetaI}
\end{align}
Hence, at linear order we can relate the curvature perturbations $\zeta$ and $\zeta^{(I)}$ by means of a simple expression involving the $\beta$ functions.

\subsection{Linearized perturbations in one field models}

In Eq. (\ref{zetadn}) we have neglected the dependence of ${\delta N}$
on the initial momenta $\pi^I_*$. Let us show more explicitly how this approximation is justified in the simple example
of a one field model. First, let us show that the full linearized solution is recovered from the $\delta N$ computation. The $e$-folding number can be written as
\begin{equation}
N= \int_{\phi_*}^{\phi_e} {H\over \dot\phi} \dd \phi.
\end{equation}
In this subsection, where we focus on one field models, we omit the index for $\phi^I$ and $\pi_I$.
The variation $\delta N$ is due to the variation of the initial value of
the field $\delta\phi_*$, and to the variation of the initial momentum
$\delta \pi_*$. The latter corresponds to a variation of the integration
constant $\Delta c_\phi$ in the solution of the H-J equation. Thus, we
have 
\begin{equation}
\delta N = -{H_* \delta\phi_* \over \dot\phi_*} + \delta N_2,
\end{equation}
where the first term is constant, and corresponds to the well known
constant solution for $\zeta$ on superhorizon scales, while the second
term is given by 
\begin{equation}
\delta N_2 = \Delta c_\phi \int_{\phi_*}^{\phi_e} {\dot\phi\over H} {\dd
 \over \dd c_\phi}\left({H\over \dot\phi}\right) \dd N. \label{seconddN}
\end{equation}
This can be cast in the form
\begin{equation}
\delta N_2=  \int \left({\Delta W\over W}-\delta_2\right)\dd N,
\end{equation}
where $\delta_2$ is defined in (\ref{deltaf2}).
Using (\ref{decay}), we have 
\begin{equation}
{\Delta W\over W} = {2\over H a^{d}} \Delta W_0.
\end{equation}
Now, by keeping track of the last term in (\ref{Def:K}), which is decaying and has been neglected in the relation (\ref{zetadn}) between $\zeta$ and ${\delta N}$, we get
\begin{equation}
\zeta= \delta N + {1\over d} \int {\partial_i\beta^i\over H} \dd
 N=\delta N-\int {\Delta W\over W} \dd N.
\end{equation} 
In the last step we have used Eq. (\ref{pibi}), and we have kept only the term of order $\epsilon^0$ in the gradient expansion, neglecting the terms of order $\epsilon^2$. Combining with Eq. (\ref{seconddN}) we obtain
\begin{equation}
\zeta_2 = -\int \delta_2  \dd N = -2d\Delta W_0 \int {c_s^2\over H
 \varepsilon_1 a^d} \dd N.
\end{equation}
This coincides with the long wavelength solution of the standard linearized equation of motion for perturbations \cite{GM}, as expected.
If the background is an attractor,  then $\delta_2$ is exponentially
decaying with $N$, and so by choosing our initial time appropriately,
$\zeta_2$ can be neglected altogether. This is the standard decaying
mode of the curvature perturbation. In the single-field case, this
confirms that we can neglect the dependence of $\delta N$ on the initial
momentum $\pi_*$.

\section{Primordial spectra from superpotential}

 \label{Sec:deltaN}

In this section we consider the spectra of the curvature and entropy perturbations, 
respectively $\zeta$ and $S^{IJ}$, at $t=t_e$ in the case of a separable product superpotential.
Let us start by considering the more elementary cross spectra for the
$\zeta^{(I)}$, which are defined in Fourier space as  
\begin{align}
  \langle \zeta^{(I)}(\bm{k}) \zeta^{(J)}(\bm{k}' ) \rangle 
& \equiv (2\pi)^d \delta(\bm{k} +
 \bm{k}') \PP_\zeta^{(I J)} (k)  \,. \label{Exp:zetap}
\end{align}
Note that ${\cal P}_{\zeta}^{(IJ)}$ is related to the power spectrum of $\delta \phi^I$ at the time $t_*$,
\begin{align}
 \langle \delta \phi^{I}_*(\bm{k})  \delta \phi^{J}_*(\bm{k}') \rangle
 \equiv (2\pi)^d \delta(\bm{k} + \bm{k}') \PP_{\phi_*}^{I J}(k)  \label{Exp:phip} \,,
\end{align}
by
\begin{align}
 \PP_\zeta^{(IJ)} (k)  
 &= N^{(I)}_{K} N^{(J)}_{L} \PP_{\phi_*}^{KL}(k) \;, \label{zetaI1I2}
\end{align}
where we have used Eq.~(\ref{Eq:deltaN}).

Likewise, we can consider higher order correlation functions. The bispectrum of $\zeta$ is defined by
\begin{align}
  \langle \zeta^{(I)}(\bm{k}_1) \zeta^{(J)}(\bm{k}_2) \zeta^{(K)}(\bm{k}_3) \rangle 
 & \equiv (2\pi)^d \delta(\bm{k}_1 + \bm{k}_2 + \bm{k}_3)
 \BB_\zeta^{(IJK)} ( k_1,\, k_2,\, k_3) \,.  \label{Exp:zetab}
\end{align}
This can be decomposed into two contributions,
\begin{equation}
\BB_{\zeta }^{(IJK)} ( k_1,\, k_2,\, k_3) = \BB_{\zeta ,\text{sub}}^{(IJK)} ( k_1,\, k_2,\, k_3) + \BB_{\zeta, \text{super}}^{(IJK)} ( k_1,\, k_2,\, k_3) \;.
\end{equation}
The first one corresponds to the intrinsic
non-Gaussianity generated until around $t=t_*$ and is related to the 3-point function of the field perturbations at horizon crossing by
\begin{align}
(2\pi)^d \delta(\bm{k}_1 + \bm{k}_2 + \bm{k}_3)
 \BB_{\zeta ,\text{sub}}^{(IJK)} ( k_1,\, k_2,\, k_3)   =   N^{(I)}_{J_1} N^{(J)}_{J_2} N^{(K)}_{J_3} \langle  \delta 
 \phi^{J_1}_*(\bm{k}_1)  \delta \phi^{J_2}_*(\bm{k}_2) \delta \phi^{J_3}_*(\bm{k}_3)
 \rangle\, .
\end{align}
The second one is generated by the super-horizon nonlinear evolution. Using Wick's theorem, it is given by
\begin{align}
   \BB_{\zeta ,\text{super}}^{(IJK)} ( k_1,\, k_2,\, k_3) 
 = N^{(I)}_{J_1} N^{(J)}_{J_2} N^{(K)}_{J_3 J_4}  \PP_{\delta \phi_*}^{J_1 J_3}(k_1) \PP_{\delta \phi_*}^{J_2 J_4}(k_2) + (2~{\rm perms})\,, \label{BI1I2I3}
\end{align}
where we have used the symmetry of $N^{(K)}_{J_3 J_4}$ with respect to the lower indices.

\subsection{Separable product superpotential}

Let us focus on the case where the Lagrangian in Eq.~\eqref{Exp:action}
is such that the metric $P_{IJ}$ is diagonal and each element $I$ only depends on the kinetic term $X^I\equiv X^{II}$ and on the field $\phi^I$,
\begin{equation}
P_{IJ} = \delta_{IJ} K_{I}(X^{I},\phi^I) \;, \qquad X^{I} = - \frac{1}{2}  \partial_\mu \phi^I \partial^\mu \phi^I  \;. \label{diag_ass}
\end{equation}
Moreover, we will assume that the superpotential is given by a separable
product, {\it i.e.},
\begin{align}
 & W(\phi^I) = \prod_{I=1}^{D} W^{(I)}(\phi^I)\,. \label{Asmp:separable}
\end{align} 
By the assumptions \eqref{diag_ass} and \eqref{Asmp:separable}, Eq.~\eqref{Eq:dphi} becomes
\begin{equation}
K_I(\phi^I,\pi_I) \frac{\dd \phi^I}{\dd N} = - \frac{d-1}{\kappa^2} \frac{\partial \ln W^{(I)}(\phi^I)}{\partial \phi^I} \;, \label{eeever2}
\end{equation}
where in the argument of $K_I$ we have traded the field velocities $\dot\phi^I$ by their 
expression
in terms of 
fields and conjugate momenta, through Eq. (\ref{fi1}) (which is also separable in this case).
Note that the momentum $\pi_I \propto \partial W/\partial\phi^I$ will actually be a function of all fields $\phi^J$. 
Hence, for the purpose of making Eq. (\ref{eeever2}) separable, we shall further restrict to the two following 
cases. 

The first case is when the kinetic term is linear in $X^{II}$ so that $K_I = K_I(\phi^I)$. In that case, we can 
always use a field redefinition to bring it to the canonical form
\begin{align}
 & P(X^I, \phi^I) =\sum_{I=1}^D X^I -V(\phi^I)\,. \label{Exp:Pcan}
\end{align}
The second case of interest is the one field case, $D=1$, in which case we can use Eq. (\ref{Eq:dphi}) to express $\pi_\phi$ as a function 
of $\phi$, for an arbitrary $P(X,\phi)$.

In these two cases, solving for the velocity and using the definition of beta function \eqref{betaup}, one obtains 
\begin{equation}
\beta^I\equiv  \kappa \frac{\dd \phi^I}{\dd N} = \beta^I(\phi^I). 
\end{equation}
Hence the evolution of $\phi^I$ as a function of $e$-folding number $N$ can be determined without being affected by the other
scalar fields. The $e$-folding number
between the slicing $\delta \phi^I=0$ at $t=t_e$ and the flat slicing at
$t=t_*$ is then given by 
\begin{align}
 & N^{(I)}(t_e,\, t_*) =
 \int^{t_e}_{t_*} H \dd t= \int^{\phi^I_e}_{\phi^I_*} \frac{\kappa \dd \phi^I}{\beta^I(\phi^I)}\,.
\end{align}
If the trajectory 
is an attractor, then after the decaying mode becomes negligibly small, 
the variation in the $e$-folding number is given by
\begin{align}
 & \delta N^{(I)} = - \frac{\kappa \delta \phi_*^I}{\beta^I_*} \,, \label{Exp:dNI}
\end{align}
with $\beta^I_* \equiv \beta^I(\phi^I_*)$. 
From this expression we can calculate the derivatives of 
$N^{(I)}$
with respect to the variations in $\delta\phi_*^J$:
\begin{align}
 N^{(I)}_J & = - \frac{\kappa}{\beta^I_*} \delta^I\!_J\,, \label{Rst:NI1} \\
 N^{(I)}_{J_1 J_2} & =\frac{\kappa}{(\beta^I_*)^2} \frac{\dd
 \beta^I_*}{\dd \phi^I_*} \delta^I\!_{J_1} \delta_{J_1 J_2} =\frac{\kappa^2}{(\beta^I_*)^3} \frac{\dd
 \beta^I_*}{\dd N_*} \delta^I\!_{J_1} \delta_{J_1 J_2}\,,  \label{Rst:NI2}
\end{align}
and so on. 
Hence, the trajectory is constructed out of $D$ independent trajectories, and there is no mixing among the different field components. 

It is known that $\delta N$ can be analytically solved also in
different examples. Under the assumption of slow-roll, it has been shown
that the number of $e$-foldings $N$ can be computed analytically when the potential
$V(\phi^I)$ is a separable product of potentials where each one depends on a single
field $\phi^I$ \cite{GBW96}, or when it is a separable sum
\cite{VW06}. This has been used to compute the non-Gaussianity in
multi-field models of inflation, in the case of slow-roll evolution in
two-field  \cite{VW06} and in multi-field models \cite{BE07, YST}. The
approach of \cite{GBW96,VW06} has been extended beyond slow-roll,
exploiting the H-J formalism in the case of a sum separable
Hubble parameter  \cite{Byrnes:2009qy, Battefeld:2009ym, Emery:2012sm}. Since by Eq.~(\ref{Eq:HW}) the Hubble
parameter $H(\phi^I)$ is nothing but the superpotential $W(\phi^I)$ (up to a
factor 2), this treatment can be straightforwardly applied to the case
of a sum separable superpotential. The approach of \cite{GBW96} was applied also to the separable
product superpotential (Hubble) case by Saffin in Ref.~\cite{Saffin}. As  
discussed in  Appendix \ref{app3}, where they overlap, our results agree with
this reference. Note that as we discussed above, for the separable product
superpotential, when the background is an attractor we can
compute $\delta N$ without introducing integration constants along the trajectory, as in
Ref.~\cite{GBW96}.  

\subsection{Primordial spectra}  \label{SSec:FormuladN}

In what follows, we shall assume that the perturbations in the different fields $\delta\phi^I_*$ are uncorrelated at
the time of horizon crossing:
\begin{align}
\PP_{\phi_*}^{IJ}(k) =\delta^{IJ} \PP_{\phi^{J}_*}(k)  \,.
 \label{Exp:Pphi}
\end{align}
In the multi-field case, Eq. (\ref{Exp:Pphi}) will hold provided
that the linearized equations of motion for perturbations decouple from each other. 
For instance, when $W^{(I)}$ are exponentials, it is easy to 
show that we can go to a basis in field space where perturbations are decoupled from each other. 
The reason is that, as we shall see, we can always change variables so that the corresponding potential $V(\phi^I)$ depends only 
on one of the fields, while the other ones are massless. In this example, the slow roll parameter $\varepsilon_1$ need not be small.
Note that for wavelengths well within the horizon, the field dependence in the potential is unimportant and perturbations of the different fields are effectively 
decoupled from each other. This suggests that in a more general setting the fields will be uncorrelated near the time of horizon crossing provided that 
$\varepsilon_2\ll 1$.

Using Eq. (\ref{Exp:Pphi}) in the
power spectrum of $\zeta^{(I)}$, Eq.~\eqref{zetaI1I2}, we have
\begin{align}
 & \PP_{\zeta}^{(IJ)}(k)  =  \delta^{KL} N^{(I)}_K N^{(J)}_L  \,
 \PP_{\phi^L_*}(k) 
 \,. 
 \label{Exp:Pk} 
\end{align}
Using Eq.~(\ref{Rst:NI1}), we obtain
\begin{align}
 & \PP^{(I J)}_{\zeta}(k) = 
 \delta^{I J}  \PP^{(I)}_{\zeta}(k) \;, \qquad \PP^{(I)}_{\zeta}(k) \equiv \frac{ \kappa^2 }{\left(\beta^{I}_*\right)^2}  \PP_{\phi^{I}_*}(k) \,, \label{Rst:Pzetas}
\end{align} 
For canonical scalar fields in
$d=3$ dimensions, the field spectrum $\PP_{\phi_*^J }$ is given by 
\begin{equation}
\PP_{\phi^{I}_*}(k)= \PP_{\phi_*}(k)= \frac{H_*^2}{2k^3} \; , 
\end{equation}
and is the same for all $I$. 

The result (\ref{Rst:Pzetas}) is also valid for the case of a single field with arbitrary speed of sound. 
In this single field case
\begin{equation}
\qquad \PP_{\phi_*}(k) = \frac{H_*^2}{2k^3 c_{s*} P_{X*}} \; , \qquad c_{s}^2 \equiv \frac{P_X}{P_X + 2 X P_{XX}} \;,
\end{equation}
where $c_{s}$
is the  speed of propagation of fluctuations  and a $*$ denotes sound-horizon crossing, $k = a_* H_*/c_{s*}$.

Defining the cross spectra for the curvature and entropy perturbations by
\begin{align}
 \langle \zeta(\bm{k}) \zeta(\bm{k}' ) \rangle 
& \equiv (2\pi)^d \delta(\bm{k} +
 \bm{k}') \PP_\zeta (k)  \,, \label{ps1}\\
 \langle \zeta(\bm{k}) S^{IJ}(\bm{k}' ) \rangle 
& \equiv (2\pi)^d \delta(\bm{k} +
 \bm{k}') \PP_{\zeta S^{IJ}} (k)  \,, \label{ps2}\\ 
 \langle S^{IJ}(\bm{k}) S^{KL}(\bm{k}' ) \rangle 
& \equiv (2\pi)^d \delta(\bm{k} +
 \bm{k}') \PP_{S^{I J}S^{KL}} (k)  \,. \label{ps3}
 \end{align}
and using Eqs.~\eqref{Def:SIJ}, (\ref{Rel:zetazetaI}) and
(\ref{Rst:Pzetas}), we obtain
\begin{align}
  \PP_{\zeta }(k) & =   \sum_{I=1}^D
 {\left( \beta_{I e} \beta^I_e  \over  \beta_e^2
 \right)^2} {\kappa^2  \over \left(\beta^I_*\right)^2} \PP_{\phi_*}\,, \label{Rst:Pzeta} \\ 
   \PP_{\zeta S^{I J}}(k)& =  d   \left[
 { \beta_{I e} \beta^{I}_e \over \beta_e^2 } \frac{\kappa^2
 }{\left(\beta^{I}_*\right)^2} -
{ \beta_{J e} \beta^{J}_e \over \beta_e^2} \frac{\kappa^2
 }{\left(\beta^{J}_*\right)^2}  
\right]\PP_{\phi_*} \,,  \label{Rst:PzetasI}  \\
  \PP_{S^{I J} S^{K L}}(k)& = d^2 
 \sum_{M=1}^D \frac{\kappa^2 }{\left(\beta^M_*\right)^2} (\delta_{I M}\! -
 \delta_{J M}\!) (\delta_{K M} - \delta_{L M})  \label{Rst:PsI} \PP_{\phi_*}\,,
\end{align}
where an index $e$ denotes a quantity evaluated
at $t=t_e$.  Thus, the only  non-vanishing components of
$\PP_{S^{IJ} S^{KL}}(k)$ are
\begin{align}
 & \PP_{S^{IJ} S^{IJ}}(k) = - \PP_{S^{IJ} S^{JI}}(k) = 
   d^2 \left[ \frac{\kappa^2}{\left(\beta^{I}_*\right)^2}  +
 \frac{\kappa^2 }{\left(\beta^{J}_*\right)^2} \right] \PP_{\phi_*}   \,.  
\end{align}
Since the power spectrum of $\zeta^I$ does not vary after the Hubble
crossing, the auto-correlation of $S^{I J}$ does not vary either.

From Eq.~\eqref{Rel:zetazetaI}, the power spectrum of $\zeta$ is given by the sum of the conserved power
spectra of $\zeta^I$ as 
\begin{align}
 & \PP_{\zeta }(k) =  \sum_{J=1}^D R_J \, \PP^{(J)}_{\zeta}(k)
 , \qquad R_J \equiv \left( { \beta_J \beta^J \over \beta^2} \right)^2
 \,.  \label{PzetawR}
\end{align}
Notice that when the trajectory converges at $t=t_e$, say in the
direction of $I=1$, satisfying
\begin{align}
 & \beta^1_e \gg \beta^I_e \sqrt{\frac{\beta^1_*}{\beta^I_*}}\, \label{Cond:single}
\end{align}
for $I \neq 1$, the amplitude of $\zeta$ is determined solely by the one
for $I=1$ at $t=t_*$ as in the single field case as
\begin{align}
 & \PP_{\zeta }(k) \simeq \PP^{(1)}_{\zeta}(k) \,.  \label{Exp:Pzetasingle}
\end{align}
Then, the influence of the components $I \neq 1$ does not explicitly
show up in the power spectrum of $\zeta$, while the change of these
components can be still traced through their contributions to
$H_*$. This property was pointed out by Garcia-Bellido and Wands in the
different separable example, where the slow-roll approximation is
employed~\cite{GBW96}.

Similarly, the bispectrum generated by the super-horizon evolution in Eq.~\eqref{BI1I2I3} can be rewritten using Eq.~\eqref{Exp:Pphi},
\begin{align}
 \BB_{\zeta ,\text{super}}^{(IJK)} ( k_1,\, k_2,\, k_3) 
  = N^{(I) }_{LM} N^{(J)}_{L} N^{(K)}_{M} \,  \PP_{\phi_*}(k_1) \PP_{\phi_* }(k_2)  + (2~{\rm perms}) \,. \label{Bispe}
\end{align}
Inserting Eqs.~(\ref{Rst:NI1}) and (\ref{Rst:NI2}) into Eq.~(\ref{Bispe}), we obtain
\begin{align}
 &\BB_{\zeta ,\text{super}}^{(IJK)} ( k_1,\, k_2,\, k_3) 
 = \delta^{IJ} \delta^{IK} {\dd \ln\beta^I_*\over \dd N_*}
 \PP^{(I)}_{\zeta}(k_1) \PP^{(I)}_{\zeta} (k_2)  + (2~{\rm perms}) \,. \label{Bispeex}
\end{align}
The bispectrum for the entropy perturbations is trivially obtained from
Eq.~(\ref{Bispeex}), since $S^{IJ}$ is linear in $\zeta^{(I)}$. In
principle, the bispectrum of $\zeta$ can also be found from
Eq.~(\ref{Bispeex}) by using the nonlinear relation between $\zeta$ and
$\zeta^{(I)}$. 

\section{$\delta N$ and holographic inflation}   \label{Sec:holography}
In Refs.~\cite{BMS, JYcsv, JKY14}, the primordial spectra were
computed holographically by means of the dual quantum field theory
which lives on the three-dimensional boundary. The computation from
holography may provide an alternative way to address the primordial
perturbations generated during inflation. In this section, we reexamine
the primordial spectra derived from the $\delta N$ formalism, comparing
them to the prediction from holography.   

\subsection{Inflation from holography}
In this subsection, we briefly overview the way to compute the
primordial perturbations from the dual boundary theory, following
Refs.~\cite{JYcsv, JKY14, Maldacena2002, JYsingle}. 
During inflation, spacetime is quasi-de Sitter, {\it i.e.},~the
de Sitter symmetry $SO(1,4)$ is slightly broken by the time evolving inflaton field. 
In order to provide the
boundary QFT which is dual to the inflationary spacetime, we need to
slightly break the conformal symmetry in $\mathbb{R}^3$, which is also
$SO(1, 4)$. In particular, to address 
a model with $D$
scalar fields, we consider a boundary QFT whose action is given by 
\begin{align}
 & S_{\rm QFT} [\chi] = S_{\rm CFT} [\chi] + \sum_{I=1}^D \int \dd \Omega_d g^I  
 {\cal O}_I(\bm{x}) \,, \label{Exp:Su}
\end{align}
where $\dd \Omega_d$ is the $d$-dimensional invariant volume and $\chi$ is
the boundary field. The second term describes the deviation from the
conformal field theory. Here, ${\cal O}_I(\bm{x})$ is a composite
operator of $\chi$ and $g^I$ are the coupling constants.
Solving the renormalization group flow, we can compute the beta
function,
\begin{align}
  & \beta_g^I \equiv \frac{\dd g^I}{\dd \ln \mu} \label{Eq:beta} \;, 
\end{align} 
as a functional of $g^I$, 
where $\mu$
is the renormalization scale. In general, since the different components
$I$ can couple with each other, solving 
Eq.~(\ref{Eq:beta}) analytically
is hard. 
However, in the case where the beta function is
separable as $\beta^I_g= \beta^I_g(g^I)$, {\it i.e.}, in the case where there is
no correlation between the different components of ${\cal O}^I$, we can
analytically solve Eq.~(\ref{Eq:beta}) as in a QFT with a single
deformation operator. In such a case, we can also compute the
auto-correlation functions of ${\cal O}(\bm{x})$ as a function of $\mu$, analytically.

Assuming that the wave functions of the curvature
and entropy perturbations are related to the generating
functional of the dual quantum fields as 
\begin{align}
 & \psi_{\rm bulk} \left[\zeta,\, s^{I'} \right] = A\, Z_{\rm QFT} \left[\zeta,\,
 s^{I'} \right]\,, \label{Exp:psi}
\end{align}
where $A$ is a normalization constant, we can compute the correlators
of $\zeta$ and $s^{I'}$. Here, to include only independent degrees of
freedom, we introduced the entropy perturbations $s^{I'}$ as 
$s^{I'} \equiv S^{1 I'}$ with $I'=2,\, \cdots,\, D$. In the boundary QFT, the primordial
perturbations $\zeta$ and $s^{I'}$ should be treated as external
fields. Once the wave function $\psi_{\rm bulk}$ is specified, we can
compute all the correlators for $\zeta$ and $s^{I'}$. For instance, the
$n$-point function for $\zeta$ is given by 
\begin{align}
 & \langle \zeta(\bm{x}_1) \zeta(\bm{x}_2)  \cdots
\zeta(\bm{x}_n) \rangle = \int  D \zeta \prod_{I'=2}^{D} D
 s^{I'} \, |\psi_{\rm bulk}|^2\, \zeta(\bm{x}_1) \zeta(\bm{x}_2)  \cdots \zeta(\bm{x}_n)\,.
\end{align}

The cosmological evolution in the bulk is described by the $D$
scalar fields as a function of  time and of the spatial
coordinates. To describe the bulk evolution by means of the dual
boundary QFT, these bulk quantities should be related to the quantities
in the boundary QFT. One may expect that the time evolution in the bulk
will be described by the RG flow in the boundary and then the time evolution
of $\phi^I$ will be determined by the RG flow of $g^I$. When we specify the
relation between the time in the bulk and the renormalization scale in
the boundary, $t=t(\mu)$, and also the relation between $\phi^I$ and
$g^I$, 
\begin{align}
 & g^I=g^I(\phi^J) \,, 
\end{align}
with the use of Eq.~(\ref{Exp:psi}), the correlators of $\zeta$ and
$s^{I'}$ can be described by $g^I$ and the correlators of ${\cal
O}$~\cite{JYcsv, JKY14}. Then, the correlators of the primordial
perturbations can be holographically computed by solving the RG flow in
the boundary.

In Refs.~\cite{BMS, AJ08, AJ09, Alex11}  it was argued that in the
de Sitter limit the renormalization scale $\mu$ should be proportional
to the scale factor $a$, 
\begin{align}
 & \mu \propto a\,. \label{Ref:mua}
\end{align}
However, the relations suggested in these references differ from each other when the solutions deviate from
de Sitter spacetime. In Ref.~\cite{JYcsv}, considering the RG flow with two fixed
points (a fixed point (FP) is a point where the beta function vanishes),
which corresponds to the time evolution in cosmology from one
de Sitter to another de Sitter, it was shown that with the choice of
Eq.~(\ref{Ref:mua}), the power spectrum of the curvature perturbation $\zeta$ in single field
models is conserved at large scales so that the holographic computation
gives a result consistent with the standard cosmological perturbation
theory. Meanwhile, a more subtle issue is left unresolved for the
conservation of the bispectrum~\cite{JYcsv}.

\subsection{Comparison of the bulk and boundary computations}    
In this subsection, we compare the result from the boundary computation,
which is obtained by solving the RG flow, to the one from the bulk
computation which is obtained in the $\delta N$ formalism. When the RG
flow with $D$ deformation operators is separable, {\it i.e.}, it is given by the $D$ copies of
the RG flow with single deformation operator, using the conformal
perturbation theory, we can derive the beta function $\beta^I_g$ as~\cite{JKY14} 
\begin{align}
 & \beta^I_g= \frac{\dd g^I(\mu)}{\dd \ln \mu} = (\Delta^I-d) g^I(\mu) +
  \frac{\pi^{d/2}}{\Gamma(d/2)} \frac{C_I}{c} \{g^I(\mu)\}^2 + {\cal O} (g^3)\,, \label{Eq:betam}
\end{align}
where $\Delta^I$ is the scaling dimension of ${\cal O}^I$, $c$ is the
central charge, and $C_I$ is the structure constant. Solving
Eq.~(\ref{Eq:betam}), we can compute $g^I(\mu)$. Once $g^I=g^I(\phi^J)$
and $t=t(\mu)$ are determined, $g^I(\mu)$ gives the time evolution
of the scalar fields $\phi^I$ in the bulk.

Assuming the relation $g^I= \kappa \phi^I$ and Eq.~(\ref{Ref:mua}) leads to $\beta_g^I=\beta^I$, where  $\beta^I$ is defined in Eq.~\eqref{beta_def}.
In this case, one can compute the primordial power spectra of $\zeta^I$  from the boundary QFT 
whose RG flow is separable. Under these assumptions, one finds \cite{JKY14}
\begin{align}
 & \PP^{(IJ)}_{\zeta} (k) = {\delta^{IJ} \over {\left[ \beta_g^I(k) \right]^2}}
 \PP_{\phi^I}(k)   \,, \label{Exp:Pzeta}
\end{align}
where $\PP_{\phi^I}(k)$ is the power spectrum of $\delta \phi^I$ in
the flat gauge,  computed from the two-point function of the
boundary operator ${\cal O}^I$. Since $\beta_g^I=\beta^I$,
this result agrees with Eq.~(\ref{Rst:Pzetas}).

Moreover,  in holography the power
spectra of $\zeta$ and $S^{IJ}$ are related to those of $\zeta^I$s by
Eqs.~(\ref{Rst:Pzeta})-(\ref{Rst:PsI}) with $\beta_g^I=\beta^I$.
Hence, the power spectra of $\zeta$ and $S^{IJ}$ computed from the
boundary QFT agree with those computed from  cosmological
perturbation theory in the bulk. 
Notice that, in the case of separable trajectories, the conserved power spectrum
of $\zeta^I$ can be computed in the same way as in single field inflation. Therefore, the fact that both calculations agree on the conservation of  
$\PP^{IJ}_{\zeta} (k)$  directly follows from the fact that they agree in the
single field case~\cite{BMS, JYcsv}.

Finally, we note that instead of using the simple relation $g^I= \kappa \phi^I$, one
may identify the couplings $g^I$ in the boundary with the scalar fields
$\phi^I$ in the bulk by more non-trivial relations $g^I=g^I(\phi^J)$. In this
case, the relation between the spectra of $\zeta^I$ and those of the
boundary operators may become more complicated, due to the Jacobian
$||\partial g^I /\partial \phi^J ||$. Yet, changing the
identification $g^I=g^I(\phi^J)$ can be simply understood  as a field
redefinition.

\section{Case studies}  \label{Sec:Case}
In this section, we compute the primordial spectra for $D$ canonical
scalar fields, using the formula derived in
Sec.~\ref{SSec:FormuladN}. For our purposes, we consider a separable
superpotential as a product of exponential superpotentials, {\it i.e.},
\begin{align}
 & W(\phi^I) = W_0 \exp\left[ - \sum_{I=1}^D f^I(\phi^I) \right]\,, \label{Exp:Wsp}
\end{align}
where $W_0$ is constant. Since the superpotential $W(\phi^I)$ directly
gives the Hubble parameter as $H=2W$, the summation of $f^I$ 
over all $I=1,\, \cdots,\, D$ should increase in time so that $H$
decreases in time.

It is instructive to consider the case where this superpotential describes $D$ scalar fields with canonical Lagrangian,
\begin{align}
 & P(X, \phi^I) = X -V(\phi^I)\;, \qquad X \equiv   - \frac{1}{2}
 \sum_{I=1}^D \partial_\mu \phi^I \partial^\mu \phi^I\,.  \label{Exp:Pcan2}
\end{align}
In this case Eqs.~(\ref{Eq:Friedmann}) and \eqref{Eq:dphi} become
\begin{align}
 & H^2 = \frac{ 2\kappa^2}{d(d-1)} \left[ X +V(\phi^I) \right]\,,  \label{Eq:Friedmann_can}
\end{align}
and 
\begin{align}
  \dot{\phi}^I = - \frac{2(d-1)}{\kappa^2} \frac{\partial
 W(\phi^I)}{\partial \phi^I} \,, \label{Eq:dphi_can} \qquad  H = 2 W(\phi^I)\,.
\end{align}
Using these equations we can compute the  potential
$V(\phi^I)$, which is given by
\begin{align}
 & V(\phi^I) = V_0
\left[1 - \frac{d-1}{d \kappa^2} \sum_{I=1}^D \ \left(  \frac{\dd {f^I}(\phi^I)}{\dd \phi^I} \right)^2  \right]  \exp\!\left[ - 2 \sum_{I=1}^D f^I(\phi^I) \right]\, , \qquad V_0 \equiv \frac{2d(d-1) W_0^2}{\kappa^2} \;.
\end{align}
Note that the potential is not, in general, separable. 
The beta function $\beta^I(\phi^I)$ is given by
\begin{align}
 & \beta^I(\phi^I) = \kappa \frac{\dd \phi^I}{\dd N} = \frac{d-1}{\kappa}  \frac{\dd {f^I}}{\dd \phi^I} \,. \label{Eq:betas}
\end{align}
In terms of  $f^I(\phi^I)$, the slow-roll parameter $\varepsilon_I$ is given by
\begin{align}
 & \varepsilon_1^I = \frac{d-1}{\kappa^2} \left(  \frac{\dd {f^I}}{\dd \phi^I} \right)^2 \,. 
 \label{Eq:eps_f}
\end{align}

\subsection{Constant $\beta^I$: Power-law inflation}
First, we consider the case where $f^I(\phi^I)$ is linear in $\phi^I$, 
\begin{align}
 & f^I(\phi^I) = \frac{1}{d-1} p_I \kappa \phi^I\,, \label{fforp}
\end{align}
where $p_I$ is a dimensionless constant parameter. Then, the potential $V(\phi^I)$
becomes a separable product of  exponential potentials for each $\phi^I$,\footnote{The case where the potential is given by a separable sum
of  exponential potentials is known as assisted inflation, see for instance~\cite{Liddle:1998jc, Malik:1998gy}.}
\begin{align}
 & V(\phi^I) = V_0
  \left(1 - \frac{1}{d(d-1)} \sum_{I=1}^{D} p_I^2  \right)\exp\left( -
 \frac{2}{d-1} \sum_{I=1}^{D} \kappa p_I \phi^I \right)\,.
\end{align}
In the single field case, this is known as power-law inflation~\cite{Lucchin:1984yf}. Using Eq.~\eqref{fforp} in Eqs.~\eqref{Eq:betas} and \eqref{Eq:eps_f} one finds that $\beta^I$ and the
slow-roll parameters $\varepsilon_I$ become constant,
\begin{align}
 & \beta^I =  p_I\,, \qquad \varepsilon_1^I =
 \frac{p_I^2}{d-1}  \,.
\end{align}
In this case, the coupling constant of the dual boundary theory $g^I$ with
$\beta^I=\beta^I_g$ blows up both at the IR and UV limits except for the
trivial case with $p_I=0$.

Solving Eq.~(\ref{Eq:betas}), we can compute the evolution of $\phi^I$
and $H$ as
\begin{align}
  \kappa \phi^I(N) &= \kappa \phi^I_* + p_I (N- N_*)   \,, \label{Sol:phipl} \\
  H(N) &  \propto  e^{ - \varepsilon_1 N } \label{Rst:Hpl} \;,
\end{align}
where $N_*$ is an integration constant and we remind the reader that $\varepsilon_1 \equiv \sum_I \varepsilon_1^I$. As  expected, integrating Eq.~(\ref{Rst:Hpl}) in time we obtain the power law
evolution
\begin{align}
 & a(t) \propto t^{1/\varepsilon_1}\,. 
\end{align}
Inflation requires $\varepsilon_1 \ll1$.

Using Eqs.~(\ref{Rst:Pzeta})--(\ref{Rst:PsI}) and assuming that all fields have the same power spectrum $\PP_{\phi_*}(k)$, the power spectra of
$\zeta$ and $S^{IJ}$  are given by
\begin{align}
  P_{\zeta \zeta}(k) & = \frac{\kappa^2}{(d-1) \varepsilon} \PP_{\phi_*}(k)
 \,, \label{Rst:Pzetapl} \\ 
  P_{\zeta S^{IJ}}(k) & =0\,, \\
   P_{S^{IJ} S^{IJ}} (k) & = \frac{(d \kappa)^2}{d-1} \left(
 \frac{1}{\varepsilon^I_{1}} + \frac{1}{\varepsilon_{1}^J} \right)\PP_{\phi_*}(k)  \;,
\end{align}
where we used $\beta^2 = (d-1) \varepsilon_1$.
The power spectrum of $\zeta$ is given by the same expression as the one
 for the single field case and the amplitude is frozen after
 $t=t_*$. For scale invariant field fluctuations in three dimensions $\PP_{\phi_*}(k) = H_*^2 /(2 k^3)$ and the spectral index is given by 
\begin{align}
 & n_s- 1 = - \frac{4 \varepsilon_1}{1 - 2 \varepsilon_1}\,. 
\end{align}
In this case, the bispectrum $B_{\zeta ,\text{super}}^{(I_1 I_2 I_3)}$ vanishes.

\subsection{Linear $\beta^I(\phi^I)$}
Next, we consider the superpotential where $f^I(\phi^I)$ also includes
the quadratic term,
\begin{align}
 & f^I(\phi^I) = \frac{1}{d-1} \left[ p_I \kappa \phi^I + q_I (\kappa \phi^I)^2 \right] \,,
\end{align}
where $p_I$ and $q_I$ are constant parameters. In this case, the beta
function $\beta^I(\phi^I)$ is given by the linear function as
\begin{align}
 &  \beta^I(\phi^I) =  p_I + 2 q_I \kappa \phi^I \,, 
\end{align}
and $V(\phi^I)$ is given by
\begin{align}
 V(\phi^I) &= V_0 \left[ 1 - \frac{1}{d(d-1)} \sum_{I=1}^{D} (p_I + 2 q_I \kappa
 \phi^I)^2 \right] 
 \exp\!\left\{ - \frac{2}{d-1}\sum_{I=1}^{D} \left[ p_I \kappa
 \phi^I + q_I (\kappa \phi^I)^2 \right] \right\}\,.
\end{align}

For $q_I \neq 0$, the potential $V(\phi^I)$ is not a separable product and
it is not easy to analytically solve the Klein-Gordon equations, which
are not separable. However, since the superpotential is a separable
product, Eq.~(\ref{Eq:betas}) can be easily solved, which gives
\begin{align}
 & \kappa \phi^I(N) =  \kappa \phi^I_* + \left( \kappa \phi^I_*  +  \frac{p_I}{2 q_I} \right) \left( e^{2 q_I (N- N_*)}  - 1 \right)  \,. \label{Sol:phiIpl}
\end{align}
As expected, in the limit $q_I \to 0$ we recover Eq.~(\ref{Sol:phipl}).
Using Eq.~(\ref{Sol:phiIpl}), we obtain the beta function $\beta^I$,
\begin{align}
 &  \beta^I(N) = \beta^I_*  \; e^{2  q_I (N -N_*)}\,.
\end{align}
At late times, the beta function blows up for $q_I > 0$ and approaches
 0 for $q_I <0$. In the perspective of holography, where the late time
in cosmology corresponds to the UV limit in the boundary QFT, the
boundary theory dual to the latter case has the FP in the UV limit $\mu \to \infty$.  
During slow-roll inflation, $\beta^I$ should be kept much smaller than 1,
requiring  
\begin{align}
 &\beta^I_* \; e^{ 2 q_I (N- N_*)} \ll 1\,. 
\end{align}

Using Eq.~\eqref{Sol:phiIpl} in Eq.~(\ref{Eq:eps_f}) we find 
\begin{align}
& \varepsilon_1^I (N)= \varepsilon_{1*}^I \; e^{ 4 q_I (N- N_*)} \;, \qquad \varepsilon_{1*}^I  \equiv  \frac{1}{d-1} \left( {p_I} + 2 q_I \kappa \phi^I_*  \right)^2 \;,
\end{align}
using which we have $f_I  = [(d-1) \varepsilon_1^I -  p_I^2]/[4 (d-1)q_I]$.
Using this in Eqs.~(\ref{Exp:Wsp}) and (\ref{Eq:dphi_can}), we can
give the following expression for the Hubble parameter
\begin{align}
 & H(N) =2 W_0  \exp \left[ \sum_I \frac{p_I^2  - (d-1) \varepsilon_1^I (N)}{4 (d-1) q_I} \right] \;.
\end{align} 
For arbitrary values of $q_I$ with $q_I \neq 0$, the Hubble parameter $H$ decreases in
time, which is simply because $\varepsilon_1>0$.

Since $\beta^I$ varies in time for $q_I \neq 0$, the power spectrum of
$\zeta$ varies also after $t=t_*$. Given that the scalar fields take
values $\phi^I_e$ at $t=t_e$ (more precisely, for each separable
trajectory of $\phi^I$, the final time $t_e$ is specified by a value of
each field, $\phi^I_e$), the power spectrum for $\zeta$ is given by the
summation of the conserved power spectra for $\zeta^I$ as in
Eq.~(\ref{PzetawR}) with the ratio $R_I$ given by
\begin{align}
 & R_I =  \frac{ \big(p_I +
 2 q_I \kappa \phi^I_e \big)^4}{\left[ \sum_{J=1}^D (p_J + 2 q_J \kappa \phi_e^J
 )^2 \right]^2}\,. 
\end{align}
Similarly, using Eqs.~(\ref{Rst:PzetasI}) and (\ref{Rst:PsI}), we can
compute the cross-correlation between $\zeta$ and $S^{IJ}$ and the
auto-correlation of $S^{IJ}$. In this case, since the beta function varies
in time, $B_{\zeta ,\text{super}}^{(I_1 I_2 I_3)}$, given in Eq.~(\ref{Bispeex}), takes a
non-vanishing value.

\subsection{Quadratic $\beta^I(\phi^I)$}
Next, we consider $f^I(\phi^I)$ which includes a cubic term as
\begin{align}
 & f^I(\phi^I) = \frac{1}{d-1} \left[ p_I \kappa \phi^I + q_I (\kappa
 \phi^I)^2 + r_I (\kappa \phi^I)^3 \right] \,,  \label{Exp:fIqd}
\end{align}
where $p_I$, $q_I$, and $r_I$ are constant parameters. Now, the beta
function $\beta^I(\phi^I)$ and the potential $V(\phi^I)$ are given by
\begin{align}
  \beta^I(\phi^I) &=  p_I + 2 q_I \kappa
 \phi^I + 3 r_I (\kappa \phi^I)^2 \,,\label{Exp:betaIqd} \\ 
  V(\phi^I) &= V_0 \left\{ 1 - \frac{1}{d(d-1)} \sum_{I=1}^{D} \left[p_I + 2 q_I \kappa
 \phi^I+ 3 r_I (\kappa \phi^I)^2 \right]^2 \right\} \cr
 & \qquad \qquad \qquad \quad  \times  \exp\!\left\{ - \frac{2}{d-1}\sum_{I=1}^{D} \left[ p_I \kappa
 \phi^I + q_I (\kappa \phi^I)^2 + r_I (\kappa \phi^I)^3 \right] \right\}\,.
\end{align}
For later use, we introduce
\begin{align}
 & D_I \equiv q_I^2 - 3r_I  p_I \,.
\end{align}
For $D_I>0$, the beta function $\beta^I$ vanishes at two
different values of $\phi^I$, {\it i.e.}, the dual boundary theory has
two FPs. For $D_I=0$, $\beta^I$ vanishes at one value of $\phi^I$, {\it
i.e.}, the boundary theory has one FP. For $D_I<0$, the beta function
$\beta^I$ does not vanish at any values of $\phi^I$, {\it i.e.},  the
boundary theory has no FPs. Using the beta function, given in
Eq.~(\ref{Exp:betaIqd}), we obtain the power spectrum of $\zeta$ as in
Eq.~(\ref{PzetawR}) with
\begin{align}
 & R_I =  \frac{ \big[p_I +
 2 q_I \kappa \phi^I_e + 3 r_I (\kappa \phi_e^I)^2 \big]^4}{\left\{ \sum_{J=1}^D
 \big[p_J + 2 q_J \kappa \phi_e^J + 3 r_J (\kappa \phi_e^J)^2 \big]^2 \right\}^2 }\,. 
\end{align}
In the following, we study the background evolution of these three cases
in turn.

\subsubsection{$D_I<0$: RG flow with no fixed point}  \label{SSec:Dn}
First, we consider the case where $\beta^I(\phi^I)$ does not
vanish. Solving Eq.~(\ref{Eq:betas}), we obtain
\begin{align}
 & \kappa \phi^I(N) = - \frac{q_I}{3 r_I} + \frac{\sqrt{-D_I}}{3 r_I} \tan \theta^I(N)  \label{Rst:phiqd}
\end{align}
with
\begin{align}
 & \theta^I(N) \equiv \sqrt{ - D_I}  (N-N_*) + \tan^{-1} \left( \frac{E_I}{\sqrt{ - D_I}} \right) \;, \qquad E_I \equiv q_I + 3 r_I \kappa \phi_*^I\,. 
\end{align}
 Inserting this solution into
Eq.~(\ref{Exp:betaIqd}), we can compute the time evolution of the beta
function $\beta^I$ as
\begin{align}
 &  \beta^I(N) =  - \frac{D_I}{3 r_I}  \left[ 1 + \tan^2 \theta^I(N) \right]\,. \label{Rst:betaIqd1}
\end{align}
The beta function starts to grow rapidly, when $\tan \theta^I(N)$ becomes
${\cal O}(1)$.
Now, the
Hubble parameter, given by 
\begin{align}
 & H(N) = H_0 \exp\left[ - \sum_{I=1}^{D}  \frac{(- D_I)^{3/2}}{27(d-1)
 r_I^2} \tan \theta^I(N) \left( \tan^2 \theta^I(N) -  3 \right) \right]
\end{align}
with
\begin{align}
 & H_0 \equiv 2 W_0 \exp\left[ \sum_{I=1}^{D} \frac{q_I}{27 (d-1) r_I^2}
 (9 r_I p_I - 2 q_I^2)  \right] \,,
\end{align}
decreases monotonically in time.

In this case, after the slow-roll time evolution, inflation ends when $\theta_I(N) \simeq \pi/4$ for at least one of the $I$s and, afterwards, 
the Hubble parameter starts to decrease more rapidly. Therefore, this case can
provide a graceful exit to inflation. Meanwhile, in the boundary side, the dual QFT does not have 
any FPs and the RG flow is dominated by irrelevant deformations in UV. It may
be interesting to study such boundary QFT.

\subsubsection{$D_I=0$: RG flow with one fixed point} 
Next, we consider the case where the beta function $\beta^I$ vanishes only at
$\kappa \phi^I= - q_I/3 r_I$. In this case, the time evolution of
$\phi^I(N)$ and $\beta^I(N)$ are given by
\begin{align}
 & \kappa \phi^I(N) = - \frac{q_I}{3 r_I} - \frac{1}{ 3 r_I (N-N_* -E_I^{-1})}\,,
\end{align}
and
\begin{align}
 & \beta^I(N) = \frac{1}{3 r_I} \frac{1}{(N-N_* - E_I^{-1})^2}\,. 
\end{align}
At late times with $(N-N_* - E_I^{-1}) \gg 1$, $\phi^I$ approaches  the constant
value $- q_I/3 r_I$, where the beta function $\beta^I$ vanishes. In this
case, the boundary theory has one FP in the UV. When all components
satisfy $D_I=0$, the Hubble parameter becomes constant at late times and the universe becomes the de Sitter spacetime.  
This solution does
not provide a realistic model of inflation because there is no graceful exit.

\subsubsection{$D_I > 0$: RG flow with two fixed points} 
Finally, we consider the case where the beta function $\beta^I$ vanishes
at two different values:
\begin{align}
 & \kappa \phi^I_{\pm} = \frac{- q_I \pm \sqrt{D_I}}{3
 r_I}\,.
\end{align}
In this case, solving Eq.~(\ref{Exp:betaIqd}), we obtain
\begin{align}
 &  \phi^I(N) = \frac{\phi_+^I + \phi_-^I e^{2 \theta^I(N)}}{1 +  e^{2 \theta^I(N)}} \,,
\end{align}
and
\begin{align}
 &  \beta^I (N) = - \frac{4 D_I}{3 r_I}
 \frac{e^{2 \theta^I(N)}}{[1+ e^{2 \theta^I(N)}]^2} \label{Rst:betaIqd3}
\end{align}
with
\begin{align}
 & \theta^I(N) \equiv \sqrt{ D_I}  (N-N_*) - \tanh^{-1} \left( \frac{E_I}{\sqrt{ D_I}} \right) \;, \qquad E_I \equiv q_I + 3 r_I \kappa \phi_*^I\,. 
\end{align}
Notice that both in the early time and late time limits, where 
$\theta \to \pm \infty$, the beta function $\beta^I(N)$ vanishes,
approaching the de Sitter spacetime. In this solution, $\phi^I$ takes
the constant values $\phi^I_-$ and $\phi^I_+$ in the limits 
$N\to - \infty$ and $N \to \infty$, respectively. If all components of
the scalar fields satisfy $D_I> 0$, the solution describes the
transition from one de Sitter to another de Sitter. In this case, the
dual boundary theory has 2 FPs both in the IR and UV and its RG flow is
driven only by relevant deformations. Such boundary theory was studied by means of the conformal
perturbation theory in Refs.~\cite{BMS, JYcsv, JKY14}

If all components satisfy $D_I \geq 0$, the universe becomes the de
Sitter spacetime at late times. Only if there exists at least one component 
$\bar I$ with $D_{\bar{I}} <0$,
 inflation can terminate as discussed in Sec.~\ref{SSec:Dn}. Notice that
the components with $D_I \geq 0$, whose $\beta^I$ decrease in time,
will satisfy 
$|\beta_{I e}| \ll |\beta_{\bar{I} e}|$
at sufficiently
late times. Then, $R_I$s become negligibly small for $I \neq \bar{I}$
and hence they do not explicitly contribute to the spectra of $\zeta$,
while they can still contribute implicitly through the Hubble parameter
and the slow-roll parameters at the Hubble crossing time.

\section{Conclusion}
\label{Sec:conclusion}

In this paper, we reviewed the superpotential formalism for multifield
inflation, and extended it to include the case of non-minimal kinetic
terms. The superpotential is useful in characterizing the attractor
behaviour of inflationary trajectories, as well as to assess the
validity of the separate universe approximation. Furthermore, the
logarithm of the superpotential plays an interesting role in the dual
description of inflation, as the c-function for the RG flow in the
boundary theory, whose gradient is related to the beta functions. 

Using the $\delta N$ formalism, we obtain simple expressions for
the power spectra for adiabatic and entropy perturbations in the case
when the superpotential is given as a separable product. In that case,
the trajectory for each field is convergent even when the whole
trajectory is not, and the power spectra can easily be found by solving
the corresponding separable background trajectories.

The bulk solution for the separable product superpotential corresponds
to a boundary QFT with a separable RG flow, where the deformation
operators are mutually uncorrelated. In such case, we showed that the
power spectra of the adiabatic and entropy perturbations computed from
the $\delta N$ formalism agree with the ones computed by solving the RG
flow of the dual boundary theory.

The separable case we addressed in this paper is described by D copies of the single field case. The power spectra of the adiabatic and
entropy perturbations can be expressed in terms of such single field
power spectra by linear relations. Because of that, the agreement of the
bulk and boundary computations for the curvature perturbation in the
single field model directly implies the agreement for the adiabatic and
entropy perturbations. It would be very  interesting to check the
agreement in more non-trivial multi field models.

Finally, with a view to phenomenological applications, we have
considered some case studies of RG flows with a polynomial c-function,
with terms up to quadratic order in the fields. These contain a range of
possible behaviours from the infrared to the UV, which may hopefully
illustrate the results which should be expected in more realistic
scenarios.

\acknowledgments
Y.~U. would like to thank C.~Byrnes, Y.~Takamizu and D.~Wands for their valuable
comments. J.~G. and Y.~U. are partially supported by MEC FPA2010-20807-C02-02 and AGAUR
2009-SGR-168. Y.~U. would like to thank University of Padua, IPhT at
CEA/Saclay, the Centro de Ciencias de Benasque Pedro Pascual, CERN, and
Institute for Advanced Study for their hospitalities during the period
of this work.  Y.~U. is supported by JSPS Grant-in-Aid for Research 
Activity Start-up under Contract No. 26887018 and the National Science
Foundation under Grant No. NSF PHY11-25915. F.V.~acknowledges partial
financial support from {\em Programme National de Cosmologie and
Galaxies} (PNCG) of CNRS/INSU, France.

\appendix

\section{Second order equations of motion from the superpotential}
\label{secondorder}
Here, we show that the usual second order equations of motion (\ref{Eq:KG}) and (\ref{Eq:Friedmann}) follow from the 
first order equation of motion (\ref{Eq:dphi}) with (\ref{Eq:HW}), for
any superpotential $W$ which satisfies the H-J equation (\ref{HJ}).  

First, we square (\ref{Eq:dphi}) to obtain
\begin{equation}
P_{IJ}P_{KL} X^{JL} = {2(d-1)^2\over \kappa^4} W_I W_K.
\end{equation}
Here, we are using the notation
\begin{equation}
W_I \equiv {\partial W\over \partial\phi^I}.\label{winot}
\end{equation}
In what follows, we assume that $P_{IJ}$ is an invertible matrix 
and we shall denote its inverse as $P^{IJ}$. We use these matrices to raise and lower the field indices. In particular
\begin{equation}
X_{IK} \equiv P_{IJ} P_{KL} X^{JL} = {2(d-1)^2 \over \kappa^2} W_IW_K,\label{lowered}
\end{equation}
and we define
\begin{equation}
X\equiv P_{IJ} X^{IJ} = P^{IJ} X_{IJ} .
\end{equation}
The H-J equation (\ref{HJ}) can then be written as
\begin{equation}
{2d(d-1)\over \kappa^2} W^2=2X-P, \label{HJS}
\end{equation}
where it is understood that any occurrence of $\dot\phi^I$ is replaced by its expression in terms of $\phi^J$ and $W_I(\phi^J)$ through Eqs. (\ref{fi1}) and (\ref{Eq:dphi}).

Taking the total derivative of (\ref{HJS}) with respect to $\phi^I$, and using (\ref{Eq:dphi}), we have
\begin{equation}
-dH P_{IJ} \dot\phi^J = 2 {\dd X\over \dd\phi^I} -{\partial P\over
 \partial \phi^I} - P_{KL} {\dd X^{KL}\over \dd\phi^I},
\end{equation}
where the $\phi^I$ dependence of $P(\phi^I,X^{IJ})$ has been separated into its explicit dependence and its dependence through field velocities contained in $X^{IJ}$.
Using
\begin{equation}
P_{KL} {\dd X^{KL}\over \dd \phi^I} = {\dd X\over \dd\phi^I} - X^{KL}
 {\dd P_{KL} \over \dd\phi^I} = {\dd X\over \dd\phi^I} + X_{KL} {\dd P^{KL} \over \dd\phi^I},
\end{equation}
and 
\begin{equation}
{\dd X\over \dd\phi^I} = P^{KL} {\dd X_{KL}\over \dd\phi^I} +
 \frac{\dd P^{KL}}{\dd \phi^I} X_{KL},
\end{equation}
we have
\begin{equation}
{\partial P\over \partial \phi^I} = dHP_{IJ}\dot\phi^J + P^{KL}{\dd X_{KL} \over \dd\phi^I}. \label{almost}
\end{equation}
We can evaluate the last term as follows:
\begin{equation}
P^{KL}{\dd X_{KL} \over \dd\phi^I}=P^{KL} {4(d-1)^2\over \kappa^4} {\partial^2 W\over \partial\phi^K\partial\phi^I} W_L 
={-2(d-1)\over \kappa^2}{\dd W_I\over \dd t} =
{\dd\over \dd t}\left(P_{IJ} \dot\phi^J\right).
\end{equation}
With this, Eq.~(\ref{almost}) coincides with the second order equation of motion for $\phi^I$, Eq.~(\ref{Eq:KG}), while the Friedmann equation (\ref{Eq:Friedmann}) is also 
satisfied because $W$ solves (\ref{HJS}).

\section{Dilution of $\partial W/\partial c_K$ with cosmic expansion}

\label{dilution}
In this appendix, we find the time dependence of the derivative of the complete solution of the H-J equation, $W(\phi^K,c_K)$,
with respect to the integration constants. From the Friedmann equation (\ref{Eq:Friedmann}), we have
\begin{equation}
{4 d(d-1) \over \kappa^2} W{\partial W\over \partial c_J} = \rho_{KL} {\partial X^{KL}\over \partial c_J}, \label{1}
\end{equation}
where we note that $\rho$ only depends on $c_K$ through the kinetic variables $X^{KL}$.
Now, from (\ref{lowered}), we have
\begin{equation}
P_{KI}P_{LJ}X^{KL}={2(d-1)^2\over\kappa^4}W_I W_J, \label{lwd}
\end{equation}
where, again, we are using the notation (\ref{winot}). Taking derivative of (\ref{lwd}) with respect to $c_K$, and then contracting with the "inverse metric" $P^{IJ}$, we immediately find
\begin{equation}
\left(2P_{IJ,LM} X^{IJ} +P_{LM}\right) {\partial X^{LM} \over
 \partial c_K} ={4(d-1)^2\over \kappa^4} P^{IJ} W_I {\partial
					   W_J\over \partial
					   c_K}. \label{eee} 
\end{equation}
Noting that $\rho= 2P_{IJ}X^{IJ}-P$, Eq.~(\ref{eee}) can be rewritten as
\begin{equation}
\rho_{LM} {\partial X^{LM} \over \partial c_K} = -2{(d-1)\over
 \kappa^2} {\dd\over \dd t} {\partial W \over \partial c_K}, \label{rhojo}
\end{equation}
where we used (\ref{Eq:dphi}) to express the right hand side as a total
time derivative. Here $\rho_{LM}$ denotes the symmetrized derivative of $\rho$
with respect to $X^{LM}$. 

Substituting Eq.~(\ref{rhojo}) in Eq.~(\ref{1}) we obtain
\begin{equation}
W{\partial W\over\partial c_J}  = - {1\over 2d}{\dd \over \dd t}{\partial W\over \partial c_J},
\end{equation}
and using $H=2W$, we have
\begin{equation}
{\dd \over \dd t}\left(\ln{\partial W\over\partial c_J}\right) = -dH,
\end{equation}
which leads to  
 \begin{equation}
{\partial W \over \partial c_J} = A^J e^{-d\int H \dd t} = A^J a^{-d},
\end{equation}
 where $A^J$ are integration constants. Hence, the dependence of $W$ on the integration constants becomes smaller 
in time along the dynamical trajectories, diluting with volume.

\section{Alternative way to compute $\delta N$} 
\label{app3}

In this appendix, we derive the primordial spectra of $\zeta$ in an
alternative way, following Refs.~\cite{GBW96, VW06, BE07}. When several
scalar fields contribute to the background evolution, the background
equations of motion can be solved analytically only for some special
cases, {\it e.g.}, the case where the background evolution can be
described by separable equations~\cite{GBW96, VW06, BE07}. In
Ref.~\cite{GBW96}, Garcia-Bellido and Wands computed the power spectrum
of $\zeta$ for the two-field model whose potential is given by a
separable product as $V(\phi_1,\, \phi_2)=V_1(\phi_1) V(\phi_2)$ under
the slow-roll assumption. In Ref.~\cite{VW06}, it was shown that a
similar analysis can be done also for a separable summation potential
$V(\phi_1,\, \phi_2) = V_1(\phi_1) + V(\phi_2)$. This analysis was
extended to the case with arbitrary number of the scalar fields~\cite{BE07} (see also Ref.~\cite{Wang:2010si}). In these
discussions, it's crucial that the Klein-Gordon equations for the scalar
fields can be recast into the first order equation by employing the
slow-roll approximation.

With the use of the
superpotential $W(\phi^I)$, which is related to the Hubble parameter as
$H(\phi^I) = 2W(\phi^I)$, the field equations for the scalar
fields can be recast into the first order equations without employing
the slow-roll approximation. In Refs.~\cite{Byrnes:2009qy,
Battefeld:2009ym}, under the assumption that the Hubble parameter is
given by the separable summation as $H(\phi^I)= \sum_{I=1}^D H_I(\phi^I)$,
Byrnes and Tasinato computed $\delta N$, following the method by Garcia-Bellido and
Wands~\cite{GBW96}. This analysis was extended to a non-canonical scalar
field in Ref.~\cite{Emery:2012sm} and to the case with the separable
summation superpotential (or the Hubble parameter) in
Ref.~\cite{Saffin}. In Appendix, we summarize the computation of $\delta
N$ in the method by Garcia-Bellido and Wands~\cite{GBW96} for the separable
product $W(\phi^I)$. The overlapped part with Ref.~\cite{Saffin} agrees.

When the superpotential $W(\phi^I)$ is given by the separable product as
in Eq.~(\ref{Asmp:separable}) and $P_{IJ}$ becomes diagonal as in
Eq.~(\ref{diag_ass}), the beta function $\beta^I$ is given by a
functional of $\phi^I$ as
\begin{align}
 & \kappa \frac{\dd \phi^I}{\dd N} = \beta^I(\phi^I)\,.  \label{Eq:betaA}
\end{align}
To solve the background trajectory, we introduce the integrals of motion $C_i$ as
\begin{align}
 & C_i \equiv \int \frac{\dd \phi^i}{\beta^i(\phi^i)} - \int \frac{\dd
 \phi^{i+1}}{\beta^{i+1}(\phi^{i+1})} \label{Def:Ci}
\end{align} 
with $i=1,\, \cdots,\, D-1$. Using Eq.~(\ref{Eq:betaA}), we can
verify that $C_i$ actually stays constant along the trajectory. With the
aid of the constant parameters $C_i$, the change of the $e$-folding along
the trajectory can be expressed only by one of the fields, say $\phi^1$, as 
\begin{align}
 & N \left(t_e,\,t_*,\, \{C_i\}_{i=1}^{{\cal N}-1} \right) =
 \int^{t_e}_{t_*} H \dd t= \int^{\phi^1_e}_{\phi^1_*} \frac{\kappa \dd
 \phi^1}{\beta^1(\phi^1)}\,.  \label{Exp:Nt}
\end{align}
On the second equality, we used $\dd t = \dd \phi^1/\dot{\phi}^1$ and
Eqs.~(\ref{Eq:HW}) and (\ref{Eq:betaA}). Unlike
$\delta N^I$, which is determined only by $\delta \phi^I_*$ for the
separable product case, the $e$-folding number (\ref{Exp:Nt}) depends
also on $\phi^I_*$ with $I \neq 1$, since $\phi^1_e$ depends on $C_i$s,
which are determined by using $\phi^I_*$ with $I=1,\, \cdots,\, D$. Now,
taking the variance with respect to $\phi^I_*$, we obtain  
\begin{align}
 & \dd N = - \frac{\kappa \dd \phi^1_*}{\beta^1_*} + \frac{1}{\beta^1_e}
 \sum_{I=1}^{{\cal N}} \frac{\partial \phi^1_e}{ \partial \phi^I_*}\, \kappa \dd
 \phi^I_*\,. \label{Exp:dNs}
\end{align}

In the following, we compute $\partial \phi^1_e/\partial \phi^I_*$. The
integrals of motion $C_i$ can be expressed in terms of $\phi^I_*$ and
hence we obtain  
\begin{align}
 & \frac{\partial \phi^I_e}{\partial \phi^J_*} = \sum_{i=1}^{D-1}
 \frac{\partial \phi^I_e}{\partial C_i} \frac{\partial C_i}{\partial
 \phi^J_*}\,, \label{Exp:dpedpsSM} 
\end{align}
Using Eq.~(\ref{Def:Ci}), we obtain
\begin{align}
 &  \frac{\partial C_i}{ \partial \phi^J_*} = \frac{1}{\beta^J_*}
 (\delta_{i\,J}- \delta_{i\,J-1})\,. \label{Exp:dCidpIsSM}
\end{align}
Choosing the uniform Hubble slicing, we specify the final time $t_e$ as
the time when the Hubble parameter takes a particular value as
\begin{align}
 & H_e = 2 W(\phi^I_e) = 2 \prod^{D}_{I=1} W^{(I)} (\phi^I_e)\,. 
\end{align}
Taking the derivative of $H_e$ with respect to $C_i$ and dividing it by
$H_e$, we obtain
\begin{align}
 & 0 = \sum_{I=1}^{D} \beta_{Ie} \, \frac{\partial \phi^I_e}{\partial
 C_i}\,. \label{Cond:HeSP}
\end{align}

Next, introducing
\begin{align}
 & \tilde{C}_I \equiv \sum_{i=1}^{I-1} C_i =  \int \frac{\dd \phi^1}{\beta^1(\phi^1)} - \int \frac{\dd
 \phi^I}{\beta^I(\phi^I)} \label{Def:tCI}
\end{align}
with $I=1,\, \cdots,\, D$, we compute 
$\partial \phi^I_e/\partial C_i$. Taking the derivative of $\tilde{C}_I$
with respect to $C_i$, we obtain
\begin{align}
 & \frac{\partial \tilde{C}_I}{\partial C_i} = \frac{1}{\beta^1_e}
 \frac{\partial \phi^1_e}{\partial C_i} - \frac{1}{\beta^I_e}
 \frac{\partial \phi^I_e}{\partial C_i}\,, \label{Exp:dtCdCSM}
\end{align}
where we noted that values of $\phi^I_e$ depend on $C_i$ chosen for each
trajectory. Equation (\ref{Exp:dtCdCSM}) is recast into 
\begin{align}
 &   \frac{\partial \phi^I_e}{\partial C_i} = \beta^I_e \left(
 \frac{1}{\beta^1_e} \frac{\partial \phi^1_e}{\partial C_i} -
 \Theta_{i\,I} \right)\,, \label{Exp:dpIedCiSM}
\end{align} 
with
\begin{eqnarray}
 & \Theta_{i\,I} \equiv {\partial \tilde{C}_I \over \partial C_i} 
 = \left\{
\begin{array}{ll}
1 & (i \leq I-1) \\
0 & (i > I-1) \\
\end{array} 
\right.
\,. 
\end{eqnarray}
Using Eqs.~(\ref{Cond:HeSP}) and (\ref{Exp:dpIedCiSM}), we obtain
\begin{align}
 &  \frac{\partial \phi^I_e}{\partial C_i} =   \beta^I_e
 \left[\frac{\sum_{J=i+1}^{D} \beta_{Ie} \beta^I_e }{ \beta_e^2} - \Theta_{iI} \right] \label{Exp:dp1eCiSM}
\end{align}
with
\begin{align}
 & \beta^2_e \equiv \sum_{I=1}^{\cal N} \beta_{Ie} \beta^I_e\,. 
\end{align}
Using Eqs.~(\ref{Exp:dCidpIsSM}) and (\ref{Exp:dp1eCiSM}), we obtain
\begin{align}
 & \frac{\partial \phi^1_e}{\partial \phi^I_*} = -
 \frac{\beta^1_e}{\beta^I_*} \left[ \frac{\beta_{Ie} \beta^I_e}{\beta^2_e} - \delta_{1I}
\right]\,, \label{Rst:dpedpsSP}
\end{align}

Inserting Eq.~(\ref{Rst:dpedpsSP}) into Eq.~(\ref{Exp:dNs}), we arrive at
the compact expression: 
\begin{align}
 & \dd N = - \sum_{I=1}^{D} \frac{\beta_{Ie} \beta^I_e}{\beta^2_e} 
 \frac{\kappa \dd \phi^I_*}{\beta^I_*}\,. \label{Res:dNs}
\end{align}
Using Eq.~(\ref{Res:dNs}), we can obtain
\begin{align}
 & N_I= \frac{\partial N}{\partial \phi^I_*} = -
 \frac{\beta_{Ie} \beta^I_e}{\beta^2_e} 
\frac{\kappa}{\beta^I_*}\,,
 \label{Rst:N1SP} \\
 & N_{IJ}= \frac{\partial^2 N}{\partial \phi^I_* \partial \phi^J_*} 
 = \delta_{IJ} \frac{\beta_{Ie} \beta^I_e}{\beta^2_e}
 \frac{\kappa^2}{(\beta^I_*)^3} \frac{\dd \beta^I_*}{\dd N_*}\,, \label{Rst:N2SP} 
\end{align}
and so on. As in the case with the separable summation 
$W(\phi^I)$, $N_{I_1 \cdots I_n}$ can be immediately computed for a
given $W_I(\phi^I)$ and take non-vanishing values only if $I_1 = \cdots =I_n$. 
The power spectrum of the curvature perturbation computed from
Eq.~(\ref{Rst:N1SP}) agrees with Eq.~(\ref{Rst:Pzeta}).



\begin{thebibliography}{99}



\bibitem{Wands:2000dp} 
  D.~Wands, K.~A.~Malik, D.~H.~Lyth and A.~R.~Liddle,
  ``A New approach to the evolution of cosmological perturbations on large scales,''
  Phys.\ Rev.\ D {\bf 62}, 043527 (2000)
  [astro-ph/0003278].
  
\bibitem{Lyth:2004gb} 
  D.~H.~Lyth, K.~A.~Malik and M.~Sasaki,
  ``A General proof of the conservation of the curvature perturbation,''
  JCAP {\bf 0505}, 004 (2005)
  [astro-ph/0411220].
  
\bibitem{Langlois:2005qp} 
  D.~Langlois and F.~Vernizzi,
  ``Conserved non-linear quantities in cosmology,''
  Phys.\ Rev.\ D {\bf 72}, 103501 (2005)
  [astro-ph/0509078].
  
\bibitem{Starobinsky:1986fxa} 
  A.~A.~Starobinsky,
  ``Multicomponent de Sitter (Inflationary) Stages and the Generation of Perturbations,''
  JETP Lett.\  {\bf 42}, 152 (1985)
  [Pisma Zh.\ Eksp.\ Teor.\ Fiz.\  {\bf 42}, 124 (1985)].


\bibitem{Salopek:1990jq} 
  D.~S.~Salopek and J.~R.~Bond,
  ``Nonlinear evolution of long wavelength metric fluctuations in inflationary models,''
  Phys.\ Rev.\ D {\bf 42}, 3936 (1990).
  

\bibitem{SS} 
  M.~Sasaki and E.~D.~Stewart,
  ``A General analytic formula for the spectral index of the density perturbations produced during inflation,''
  Prog.\ Theor.\ Phys.\  {\bf 95}, 71 (1996)
  [astro-ph/9507001].

\bibitem{Sasaki:1998ug} 
  M.~Sasaki and T.~Tanaka,
  ``Superhorizon scale dynamics of multiscalar inflation,''
  Prog.\ Theor.\ Phys.\  {\bf 99}, 763 (1998)
  [gr-qc/9801017].

\bibitem{LMS} 
  D.~H.~Lyth, K.~A.~Malik and M.~Sasaki,
  ``A General proof of the conservation of the curvature perturbation,''
  JCAP {\bf 0505}, 004 (2005)
  [astro-ph/0411220].



\bibitem{LR98} 
  D.~H.~Lyth and A.~Riotto,
  ``Particle physics models of inflation and the cosmological density perturbation,''
  Phys.\ Rept.\  {\bf 314}, 1 (1999)
  [hep-ph/9807278].


\bibitem{Tanaka:2010km} 
  T.~Tanaka, T.~Suyama and S.~Yokoyama,
  ``Use of delta N formalism - Difficulties in generating large local-type non-Gaussianity during inflation -,''
  Class.\ Quant.\ Grav.\  {\bf 27}, 124003 (2010)
  [arXiv:1003.5057 [astro-ph.CO]].

\bibitem{Sasaki:2006kq} 
  M.~Sasaki, J.~Valiviita and D.~Wands,
  ``Non-Gaussianity of the primordial perturbation in the curvaton model,''
  Phys.\ Rev.\ D {\bf 74}, 103003 (2006)
  [astro-ph/0607627].

\bibitem{LVW08} 
  D.~Langlois, F.~Vernizzi and D.~Wands,
  ``Non-linear isocurvature perturbations and non-Gaussianities,''
  JCAP {\bf 0812}, 004 (2008)
  [arXiv:0809.4646 [astro-ph]].

\bibitem{Skenderis:2006jq} 
  K.~Skenderis and P.~K.~Townsend,
  ``Hidden supersymmetry of domain walls and cosmologies,''
  Phys.\ Rev.\ Lett.\  {\bf 96}, 191301 (2006)
  [hep-th/0602260].


\bibitem{Skenderis:2006fb} 
  K.~Skenderis and P.~K.~Townsend,
  ``Pseudo-Supersymmetry and the Domain-Wall/Cosmology Correspondence,''
  J.\ Phys.\ A {\bf 40}, 6733 (2007)
  [hep-th/0610253].

\bibitem{Larsen:2002et} 
  F.~Larsen, J.~P.~van der Schaar and R.~G.~Leigh,
  ``De Sitter holography and the cosmic microwave background,''
  JHEP {\bf 0204}, 047 (2002)
  [hep-th/0202127].

\bibitem{LM03}
  F.~Larsen and R.~McNees,
  ``Inflation and de Sitter holography,''
  JHEP {\bf 0307}, 051 (2003)
  [hep-th/0307026].

\bibitem{vdS}
  J.~P.~van der Schaar,
  ``Inflationary perturbations from deformed CFT,''
  JHEP {\bf 0401}, 070 (2004)
  [hep-th/0307271].

\bibitem{LM04}
  F.~Larsen and R.~McNees,
  ``Holography, diffeomorphisms, and scaling violations in the CMB,''
  JHEP {\bf 0407}, 062 (2004)
  [hep-th/0402050].





\bibitem{Strominger}
  A.~Strominger,
  ``The dS / CFT correspondence,''
  JHEP {\bf 0110}, 034 (2001)
  [hep-th/0106113].


\bibitem{Witten}
  E.~Witten,
  ``Quantum gravity in de Sitter space,''
  hep-th/0106109.



\bibitem{Bousso:2001mw}
  R.~Bousso, A.~Maloney and A.~Strominger,
  ``Conformal vacua and entropy in de Sitter space,''
  Phys.\ Rev.\ D {\bf 65}, 104039 (2002)
  [hep-th/0112218].



\bibitem{Strominger2}
  A.~Strominger,
  ``Inflation and the dS / CFT correspondence,''
  JHEP {\bf 0111}, 049 (2001)
  [hep-th/0110087].


\bibitem{Anninos:2011ui}
  D.~Anninos, T.~Hartman and A.~Strominger,
  ``Higher Spin Realization of the dS/CFT Correspondence,''
  arXiv:1108.5735 [hep-th].
 
  
\bibitem{BMS} 
  A.~Bzowski, P.~McFadden and K.~Skenderis,
  ``Holography for inflation using conformal perturbation theory,''
  JHEP {\bf 1304}, 047 (2013)
  [arXiv:1211.4550 [hep-th]].


\bibitem{JYcsv}
  J.~Garriga and Y.~Urakawa,
  ``Holographic inflation and the conservation of $\zeta$,''
  JHEP {\bf 1406}, 086 (2014)
  [arXiv:1403.5497 [hep-th]].

\bibitem{JKY14}
  J.~Garriga, K.~Skenderis and Y.~Urakawa,
  ``Multi-field inflation from holography,''
  JCAP {\bf 1501}, no. 01, 028 (2015)
  [arXiv:1410.3290 [hep-th]].



\bibitem{Maldacena2002}
  J.~M.~Maldacena,
  ``Non-Gaussian features of primordial fluctuations in single field inflationary models,''
  JHEP {\bf 0305} (2003) 013
  [astro-ph/0210603].



\bibitem{Seery:2006tq}
  D.~Seery and J.~E.~Lidsey,
  ``Non-Gaussian Inflationary Perturbations from the dS/CFT Correspondence,''
  JCAP {\bf 0606}, 001 (2006)
  [astro-ph/0604209].


 
\bibitem{Maldacena:2011nz} 
  J.~M.~Maldacena and G.~L.~Pimentel,
  ``On graviton non-Gaussianities during inflation,''
  JHEP {\bf 1109}, 045 (2011)
  [arXiv:1104.2846 [hep-th]].
  


\bibitem{Schalm:2012pi} 
  K.~Schalm, G.~Shiu and T.~van der Aalst,
  ``Consistency condition for inflation from (broken) conformal symmetry,''
  JCAP {\bf 1303}, 005 (2013)
  [arXiv:1211.2157 [hep-th]].

\bibitem{Mata:2012bx}
  I.~Mata, S.~Raju and S.~Trivedi,
  ``CMB from CFT,''
  JHEP {\bf 1307}, 015 (2013)
  [arXiv:1211.5482 [hep-th]].

\bibitem{JYsingle}
  J.~Garriga and Y.~Urakawa,
  ``Inflation and deformation of conformal field theory,''
  JCAP {\bf 1307}, 033 (2013)
  [arXiv:1303.5997 [hep-th]].



\bibitem{Ghosh:2014kba} 
  A.~Ghosh, N.~Kundu, S.~Raju and S.~P.~Trivedi,
  ``Conformal Invariance and the Four Point Scalar Correlator in Slow-Roll Inflation,''
  JHEP {\bf 1407}, 011 (2014)
  [arXiv:1401.1426 [hep-th]].
  
  
  
\bibitem{Larsen:2014wpa}
  F.~Larsen and A.~Strominger,
  ``BICEP2 and the Central Charge of Holographic Inflation,''
  arXiv:1405.1762 [hep-th].


\bibitem{Banks:2013qra}
  T.~Banks, W.~Fischler, T.~J.~Torres and C.~L.~Wainwright,
  ``Holographic Fluctuations from Unitary de Sitter Invariant Field Theory,''
  arXiv:1306.3999 [hep-th].

\bibitem{Banks:2013qpa}
  T.~Banks,
  ``Lectures on Holographic Space Time,''
  arXiv:1311.0755 [hep-th].

\bibitem{Kiritsis:2013gia}
  E.~Kiritsis,
  ``Asymptotic freedom, asymptotic flatness  and cosmology,''
  JCAP {\bf 1311}, 011 (2013)
  [arXiv:1307.5873 [hep-th]].




\bibitem{Kol:2013msa}
  U.~Kol,
  ``On the dual flow of slow-roll Inflation,''
  JHEP {\bf 1401}, 017 (2014)
  [arXiv:1309.7344 [hep-th]].

\bibitem{Kawai:2014vxa} 
  S.~Kawai and Y.~Nakayama,
  ``Improvement of energy-momentum tensor and non-Gaussianities in holographic cosmology,''
  JHEP {\bf 1406}, 052 (2014)
  [arXiv:1403.6220 [hep-th]].



\bibitem{Binetruy:2014zya} 
  P.~Binetruy, E.~Kiritsis, J.~Mabillard, M.~Pieroni and C.~Rosset,
  ``Universality classes for models of inflation,''
  JCAP {\bf 1504}, no. 04, 033 (2015)
  [arXiv:1407.0820 [astro-ph.CO]].





\bibitem{MS_HC09}
  P.~McFadden and K.~Skenderis,
  ``Holography for Cosmology,''
  Phys.\ Rev.\  D {\bf 81}, 021301 (2010)
  [arXiv:0907.5542 [hep-th]].



\bibitem{MS_HC10}
  P.~McFadden, K.~Skenderis,
  ``The Holographic Universe,''
  J.\ Phys.\ Conf.\ Ser.\  {\bf 222}, 012007 (2010).
  [arXiv:1001.2007 [hep-th]].

\bibitem{MS_HCob10}
  P.~McFadden, K.~Skenderis,
  ``Observational signatures of holographic models of inflation,''
  [arXiv:1010.0244 [hep-th]].

\bibitem{MS_NG}
  P.~McFadden, K.~Skenderis,
  ``Holographic Non-Gaussianity,''
  JCAP {\bf 1105}, 013 (2011).
  [arXiv:1011.0452 [hep-th]].

\bibitem{MS_NGGW}
  P.~McFadden and K.~Skenderis,
  ``Cosmological 3-point correlators from holography,''
  JCAP {\bf 1106}, 030 (2011)
  [arXiv:1104.3894 [hep-th]].

 \bibitem{McFadden:2013ria} 
  P.~McFadden,
  ``On the power spectrum of inflationary cosmologies dual to a deformed CFT,''
  JHEP {\bf 1310}, 071 (2013)
  [arXiv:1308.0331 [hep-th]].

\bibitem{Seery:2008ax} 
  D.~Seery, M.~S.~Sloth and F.~Vernizzi,
  ``Inflationary trispectrum from graviton exchange,''
  JCAP {\bf 0903}, 018 (2009)
  [arXiv:0811.3934 [astro-ph]].



\bibitem{Skenderis:1999mm}
  K.~Skenderis and P.~K.~Townsend,
  ``Gravitational stability and renormalization group flow,''
  Phys.\ Lett.\ B {\bf 468} (1999) 46
  [hep-th/9909070].

\bibitem{DeWolfe:1999cp} 
  O.~DeWolfe, D.~Z.~Freedman, S.~S.~Gubser and A.~Karch,
  ``Modeling the fifth-dimension with scalars and gravity,''
  Phys.\ Rev.\ D {\bf 62}, 046008 (2000)
  [hep-th/9909134].


\bibitem{Freedman:2003ax} 
  D.~Z.~Freedman, C.~Nunez, M.~Schnabl and K.~Skenderis,
  ``Fake supergravity and domain wall stability,''
  Phys.\ Rev.\ D {\bf 69}, 104027 (2004)
  [hep-th/0312055].




\bibitem{Kinney:1997ne} 
  W.~H.~Kinney,
  ``A Hamilton-Jacobi approach to nonslow roll inflation,''
  Phys.\ Rev.\ D {\bf 56}, 2002 (1997)
  [hep-ph/9702427].



  \bibitem{zamo} 
  A.~B.~Zamolodchikov,
  JETP Lett.\  {\bf 43}, 730 (1986)
  [Pisma Zh.\ Eksp.\ Teor.\ Fiz.\  {\bf 43}, 565 (1986)].


\bibitem{TDSD}
  D.~Langlois, S.~Renaux-Petel, D.~A.~Steer and T.~Tanaka,
  ``Primordial perturbations and non-Gaussianities in DBI and general multi-field inflation,''
  Phys.\ Rev.\ D {\bf 78} (2008) 063523
  [arXiv:0806.0336 [hep-th]].

\bibitem{GM} 
  J.~Garriga and V.~F.~Mukhanov,
  ``Perturbations in k-inflation,''
  Phys.\ Lett.\ B {\bf 458}, 219 (1999)
  [hep-th/9904176].


\bibitem{SKF} 
  N.~S.~Sugiyama, E.~Komatsu and T.~Futamase,
  ``$\delta$N formalism,''
  Phys.\ Rev.\ D {\bf 87}, no. 2, 023530 (2013)
  [arXiv:1208.1073 [gr-qc]].

\bibitem{NTS} 
  A.~Naruko, Y.~i.~Takamizu and M.~Sasaki,
  ``Beyond $\delta N$ formalism,''
  PTEP {\bf 2013}, 043E01 (2013)
  [arXiv:1210.6525 [astro-ph.CO]].



\bibitem{Langlois:2011zz} 
  D.~Langlois and A.~Lepidi,
  ``General treatment of isocurvature perturbations and non-Gaussianities,''
  JCAP {\bf 1101}, 008 (2011)
  [arXiv:1007.5498 [astro-ph.CO]].



\bibitem{GBW96} 
  J.~Garcia-Bellido and D.~Wands,
  ``Metric perturbations in two field inflation,''
  Phys.\ Rev.\ D {\bf 53}, 5437 (1996)
  [astro-ph/9511029].




\bibitem{VW06} 
  F.~Vernizzi and D.~Wands,
  ``Non-gaussianities in two-field inflation,''
  JCAP {\bf 0605}, 019 (2006)
  [astro-ph/0603799].


\bibitem{BE07} 
  T.~Battefeld and R.~Easther,
  ``Non-Gaussianities in Multi-field Inflation,''
  JCAP {\bf 0703}, 020 (2007)
  [astro-ph/0610296].

\bibitem{YST} 
  S.~Yokoyama, T.~Suyama and T.~Tanaka,
  ``Primordial Non-Gaussianity in Multi-Scalar Inflation,''
  Phys.\ Rev.\ D {\bf 77}, 083511 (2008)
  [arXiv:0711.2920 [astro-ph]].


\bibitem{Byrnes:2009qy} 
  C.~T.~Byrnes and G.~Tasinato,
  ``Non-Gaussianity beyond slow roll in multi-field inflation,''
  JCAP {\bf 0908}, 016 (2009)
  [arXiv:0906.0767 [astro-ph.CO]].

\bibitem{Battefeld:2009ym} 
  D.~Battefeld and T.~Battefeld,
  ``On Non-Gaussianities in Multi-Field Inflation (N fields): Bi and Tri-spectra beyond Slow-Roll,''
  JCAP {\bf 0911}, 010 (2009)
  [arXiv:0908.4269 [hep-th]].

\bibitem{Emery:2012sm} 
  J.~Emery, G.~Tasinato and D.~Wands,
  ``Local non-Gaussianity from rapidly varying sound speeds,''
  JCAP {\bf 1208}, 005 (2012)
  [arXiv:1203.6625 [hep-th]].

\bibitem{Saffin} 
  P.~M.~Saffin,
  ``The covariance of multi-field perturbations, pseudo-susy and $f_{NL}$,''
  JCAP {\bf 1209}, 002 (2012)
  [arXiv:1203.0397 [hep-th]].


\bibitem{AJ08} 
  J.~Garriga and A.~Vilenkin,
  ``Holographic Multiverse,''
  JCAP {\bf 0901}, 021 (2009)
  [arXiv:0809.4257 [hep-th]].
  
  
\bibitem{AJ09} 
  J.~Garriga and A.~Vilenkin,
  ``Holographic multiverse and conformal invariance,''
  JCAP {\bf 0911}, 020 (2009)
  [arXiv:0905.1509 [hep-th]].



  
\bibitem{Alex11} 
  A.~Vilenkin,
  ``Holographic multiverse and the measure problem,''
  JCAP {\bf 1106}, 032 (2011)
  [arXiv:1103.1132 [hep-th]].


\bibitem{Liddle:1998jc} 
  A.~R.~Liddle, A.~Mazumdar and F.~E.~Schunck,
  ``Assisted inflation,''
  Phys.\ Rev.\ D {\bf 58}, 061301 (1998)
  [astro-ph/9804177].


\bibitem{Malik:1998gy} 
  K.~A.~Malik and D.~Wands,
  ``Dynamics of assisted inflation,''
  Phys.\ Rev.\ D {\bf 59}, 123501 (1999)
  [astro-ph/9812204].



\bibitem{Lucchin:1984yf} 
  F.~Lucchin and S.~Matarrese,
  ``Power Law Inflation,''
  Phys.\ Rev.\ D {\bf 32}, 1316 (1985).













\bibitem{Wang:2010si} 
  T.~Wang,
  ``Note on Non-Gaussianities in Two-field Inflation,''
  Phys.\ Rev.\ D {\bf 82}, 123515 (2010)
  [arXiv:1008.3198 [astro-ph.CO]].


\end{thebibliography}
\end{document}